%% file: aps-9.tex
\documentclass[pre,aps,twocolumn]{revtex4-1}

\pdfoutput=1

\usepackage{bm}
\usepackage{epsfig}
\usepackage{amsmath}
\usepackage{graphicx}
\usepackage{amssymb}
\usepackage[normalem]{ulem}
\usepackage[usenames]{color}

\begin{document}

\newcommand{\kB}{k_\mathrm{B}}
\newcommand{\tw}{t_\mathrm{w}}
\newcommand{\ee}{\mathrm{e}}

\definecolor{Blue}{rgb}{0,0.0,1.0}
\newcommand{\comment}[1]{\textcolor{Blue}{#1}}

\title{Predicting the self-assembly of a model colloidal crystal}
\author{Daphne Klotsa}
\author{Robert L. Jack}
\affiliation{Department of Physics, University of Bath, Bath BA2 7AY, United Kingdom}

\begin{abstract}
We investigate the self-assembly (crystallisation) of 
particles with hard cores and isotropic, square-well interactions, using a Monte Carlo scheme to
simulate overdamped Langevin dynamics.  We measure correlation and response functions
during the early stages of assembly, and we analyse the results using fluctuation-dissipation theorems,
aiming to predict 
which systems will self-assemble successfully and which will get stuck in disordered states.  The early-time
correlation and response
measurements are made before significant crystallisation has taken place, 
indicating that dynamical measurements are valuable in measuring a system's
propensity for kinetic trapping.
\end{abstract}

\maketitle

\section{Introduction}

\input{body-9.tex}

\begin{acknowledgments}
We thank Mike Hagan, Paddy Royall, Steve Whitelam, and David Chandler for many useful discussions.  This
work was supported by the EPSRC through grants EP/G038074/1 and EP/I003797/1.
\end{acknowledgments}

\begin{appendix}

\input{appendices-v3.tex}

\end{appendix}

\bibliography{myrsc} 
\bibliographystyle{apsrev4-1} 

\end{document}

%% file: body-9.tex
The self-assembly of individual components into ordered structures has been studied in a variety of contexts including biomaterials research~\cite{langer04, stupp10} and nanoengineering~\cite{love05, mastra09, white09, sri10, wang10, grzel10,sol07}. Significant progress has been made in experimentally synthesising different building blocks designed to assemble into specific target products~\cite{hong06,doug09,sac10}.  Assembling systems, however, often get stuck in metastable disordered states before reaching the equilibrium ordered ones. Kinetic traps are absent from classical theories of phase change~\cite{bray} but they pose real problems in experiments such as those on protein and colloidal crystallisation \cite{slab07,leng09,lu06}. In order to address the kinetics of self-assembly, out-of-equilibrium theories need to be developed.

Crystallisation is a well-studied process, whereby particles in a fluid state undergo a phase transition and develop long-ranged crystalline order.  
This spontaneous ordering is an example of self-assembly.  
Controlling the crystallisation of proteins and colloids would have applications in biology (for example, in determining the structure
of biomolecules~\cite{Chayen2004})
as well as in photonics (for example, in controlling the propagation of light~\cite{vlasov2001,hynn2007}). 
In both colloidal and certain biomolecular systems the constituent particles interact via an effective short-ranged attractive potential with repulsive 
`hard' cores~\cite{frenkel97,likos01,foffi02,dijkstra02,tava04,charb07,fortini08}. 

\begin{figure}
\centering
\includegraphics[width=8.5cm]{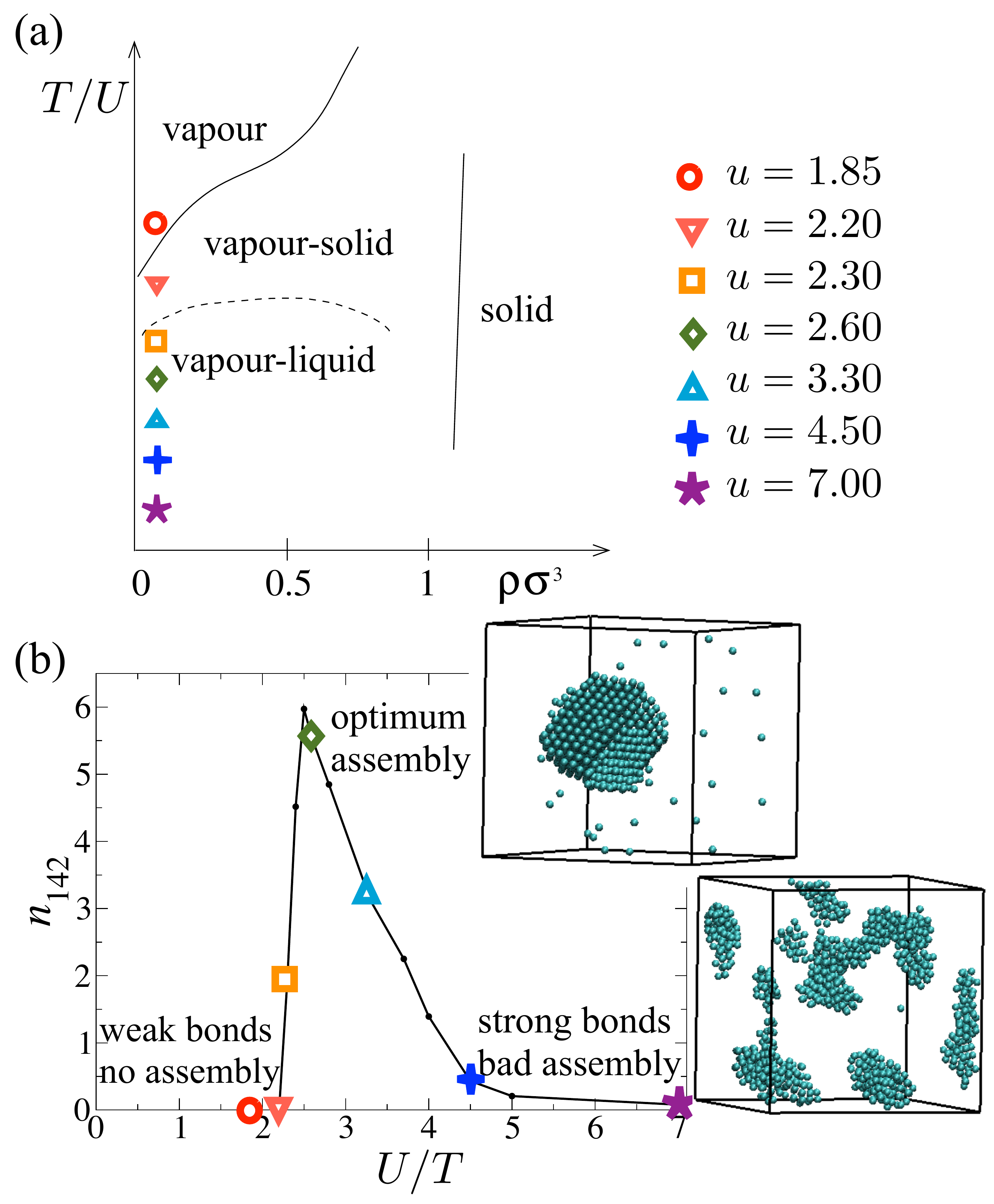}
\caption{(a) Schematic phase diagram for a square-well potential, following Ref.~\cite{liu05}. The legend shows the 
bond strengths $U$ that will be considered in this work, and we show estimates of the locations of these state points
within the schematic phase diagram. 
(b)~We plot a measure of the crystallinity of the system, for various bond strengths, at a time $t=10^6$ MC steps. The crystallinity is measured using a common neighbour analysis~\cite{hon87}, as described in the main text. The colours of the symbols indicate the transition from high temperatures (red, weak bonds) towards low temperatures (violet, strong bonds).}
\label{fig:phase-yield}
\end{figure}

In this paper, we study a dilute suspension of attractive colloidal particles that separate into crystal and fluid phases, interpreting crystallisation as a simple example of self-assembly, as in Ref.~\cite{whitelam09}. A schematic phase diagram is shown in Fig.~\ref{fig:phase-yield}(a), with points indicating seven characteristic bond strengths that we have considered. The systems we investigate begin in a far-from equilibrium fluid state, and evolve towards the stable crystal: we are interested in the \emph{dynamics} of this process. The questions we address are the following: what happens during the process of assembly? How does a system that will eventually assemble into a crystal differ from one that will get kinetically trapped? What are the first signs of frustration and when do they appear? How does the strength of attractive interactions influence the dynamics? What are the relevant physical quantities one should measure in order to understand the dynamics, and at what times should we measure them? 

Self-assembly processes require a balance between a net drive to assembly and kinetic accessibility. The first requirement implies that the thermodynamic equilibrium state is an assembled one, which presupposes sufficiently strong attraction between individual components. The second requirement is kinetic and ensures that the thermodynamic state is accessible within reasonable timescales. The two requirements are competing, with the thermodynamic one favouring low temperatures and strong bonds and the kinetic one high temperatures and weaker bonds. Fig.~\ref{fig:phase-yield}(b) shows the crystallinity of the system as a function of the bond strength: its non-monotonic form demonstrates the competition between kinetic and thermodynamic effects, with a clear maximum at intermediate bond strengths.

In the past, the two competing requirements for effective self-assembly have been characterised through fluctuation-dissipation ratios (FDRs), which measure the extent to which equilibrium fluctuation-dissipation theorems (FDTs) continue to hold in out-of-equilibrium settings. In particular, it was shown~\cite{jack07} that correlation and response functions can provide a measure of the balance between the net drive to assembly and kinetic accessibility, and therefore indicate which systems will assemble into ordered structures and which ones will get kinetically trapped. Jack, Hagan and Chandler~\cite{jack07} investigated and compared two models: viral capsids and sticky disks. 
The purpose of this paper is to extend these ideas into a model colloidal system in order 
(i) to study whether the fate of this system can be predicted from the early stages of assembly based on FDR measurements, thus providing insight and subsequently guidelines for colloidal and protein crystallisation experiments and simulations; 
and (ii) to identify generic features of self-assembly that are not system-dependent.

Details of the computational model are given in section~\ref{sec:model}. In Section~\ref{sec:cryst} we show how the system evolves in time for different interaction strengths, identifying regimes of slow nucleation, rapid assembly and kinetic trapping -- features that are not predicted by the equilibrium phase diagram. We then introduce correlation and response functions in Section~\ref{sec:predict}, explaining how they can be used to predict assembly quality. We include a discussion of the robustness of our measurements: how they depend on measurement times, interaction range and volume fraction. Finally, Section~\ref{sec:conc} gives a discussion of the results and poses questions for future investigation. 

\section{Model}
\label{sec:model}

We study a system of $N$ spherical particles in a cubic box of volume $V$, with periodic boundary conditions.  The particles have hard cores and isotropic, square-well interactions. Their hard-core diameter is $\sigma$, the depth of the potential is $U$ and the range of interaction is $\xi \sigma$. Thus,
the energy of the system is
\begin{equation}
E = -\frac{U}2 \sum_i n_i 
\end{equation}
where $n_i$ is the number of neighbours for particle $i$: that is, the number of particles within the interaction range $\xi \sigma$.
It is also convenient to define $E_i=-\frac12 n_i U$.

This model is a simple but effective description for colloids and some globular proteins~\cite{ash96,liu05,duda09}.  
As shown for example in Liu \emph{et al.}~\cite{liu05}, the short range of the attractions reduces the stability of the liquid phase, and the vapour-liquid binodal is metastable, lying within the vapour-solid coexistence curve. The phase diagram
is sketched in Fig.~\ref{fig:phase-yield}(a), where we have labelled the state points that we focus on in this article.
The corresponding bond strengths are shown in the legend; for simplicity we have defined $u=U/T$ which will be used to quote the bond strengths in the rest of the paper (we take Boltzmann's constant $k_{\rm{B}}=1$ throughout). The colours of the symbols indicate the transition from high temperatures (red circles) to low temperatures (violet stars), in an order that follows the spectrum of visible light.

We have carried out dynamical simulations of this model.
Beginning with equilibrated systems of hard spheres, we use a Monte Carlo (MC) scheme as an approximate method to simulate the overdamped Langevin 
equation
\begin{equation}
\frac{\partial}{\partial t}\bm{r}_i = - \frac{D}{\kB T} \nabla_i E + \bm{\eta}_i
\label{equ:lang}
\end{equation}
where $E$ is the total energy of the system,
$\nabla_i=(\frac\partial{\partial x_i},\frac\partial{\partial y_i},\frac\partial{\partial z_i})$ 
as usual\footnote{
  Strictly, $\nabla_i E$ is ill-defined, since $E$ is not a continuous function of the particle coordinates.  For the purposes
  of (\ref{equ:lang}) we imagine regularising the 
  square-well potential by taking a limiting case of a smooth but steep potential.  In practice, we integrate this equation using a Monte Carlo 
  scheme with a finite time step $\tau_0$, which avoids the need for any explicit regularisation.
},
and the components of the vectors $\bm{\eta}_i$ are independent white noises. 

On each move of our Monte Carlo scheme, we pick a random particle and propose a random displacement from a cube of side $2a_0$, centred at the origin.  (Thus, the maximum displacement in each of the $x$, $y$ and $z$ directions
is $a_0$.)  We accept the move with probability $\min(1,\ee^{-\Delta E/\kB T})$ where $\Delta E$ is the energy difference between the states before and after the proposed move.
 A Monte Carlo step (or sweep) consists of $N$ Monte Carlo moves, and we associate it with a time increment $\tau_0$.
In the limit of $a_0,\tau_0\to0$ while holding $D=a_0^2/6\tau_0$ constant, this method provides dynamical trajectories in accordance
with the Langevin equation~(\ref{equ:lang}) above\footnote{
  An alternative to our MC scheme would be to use Brownian dynamics to simulate this system: in the limit of small time step $\tau_0$ then
  both Brownian and MC dynamics are equivalent.  One reason to prefer the MC in this study is that the fluctuation-dissipation 
  theorems described in the following sections hold exactly for equilibrated systems with MC dynamics, even when the time step $\tau_0$ is finite.
  (This is not the case when using Brownian dynamics.)
}.
In the Langevin description, the natural unit of time is the Brownian time $\tau_\mathrm{B}=\sigma^2/D=(\sigma/a_0)^2\tau_0$.
The behaviour of the model depends on the dimensionless parameters $\xi$ and $u=U/T$ as well as on the particle volume fraction.
Unless otherwise stated, our simulations are done at volume fraction $4\%$ (i.e., $\pi N\sigma^3/6V=0.04$), with $N$=1000 and $\xi$=0.11$\sigma$, and we take $a_0=0.15\sigma$. 
We quote times in MC steps (units of $\tau_0$), noting that $\tau_\mathrm{B}\approx44\tau_0$ for our chosen
step size $a_0$.
We observe that this step size $a_0$ is comparable to the interaction range $\xi\sigma$, so that our results are not yet representative of the limit of small $a_0$.  We have conducted simulations with smaller $a_0$: while quantitative differences are observed, qualitative features are unchanged.  As usual, our time step (or equivalently, $a_0$) is chosen as a trade-off between accuracy of numerical integration and practical efficiency.

\section{Crystallisation process}
\label{sec:cryst}

\begin{figure}
\centering
\includegraphics[width=7.0cm]{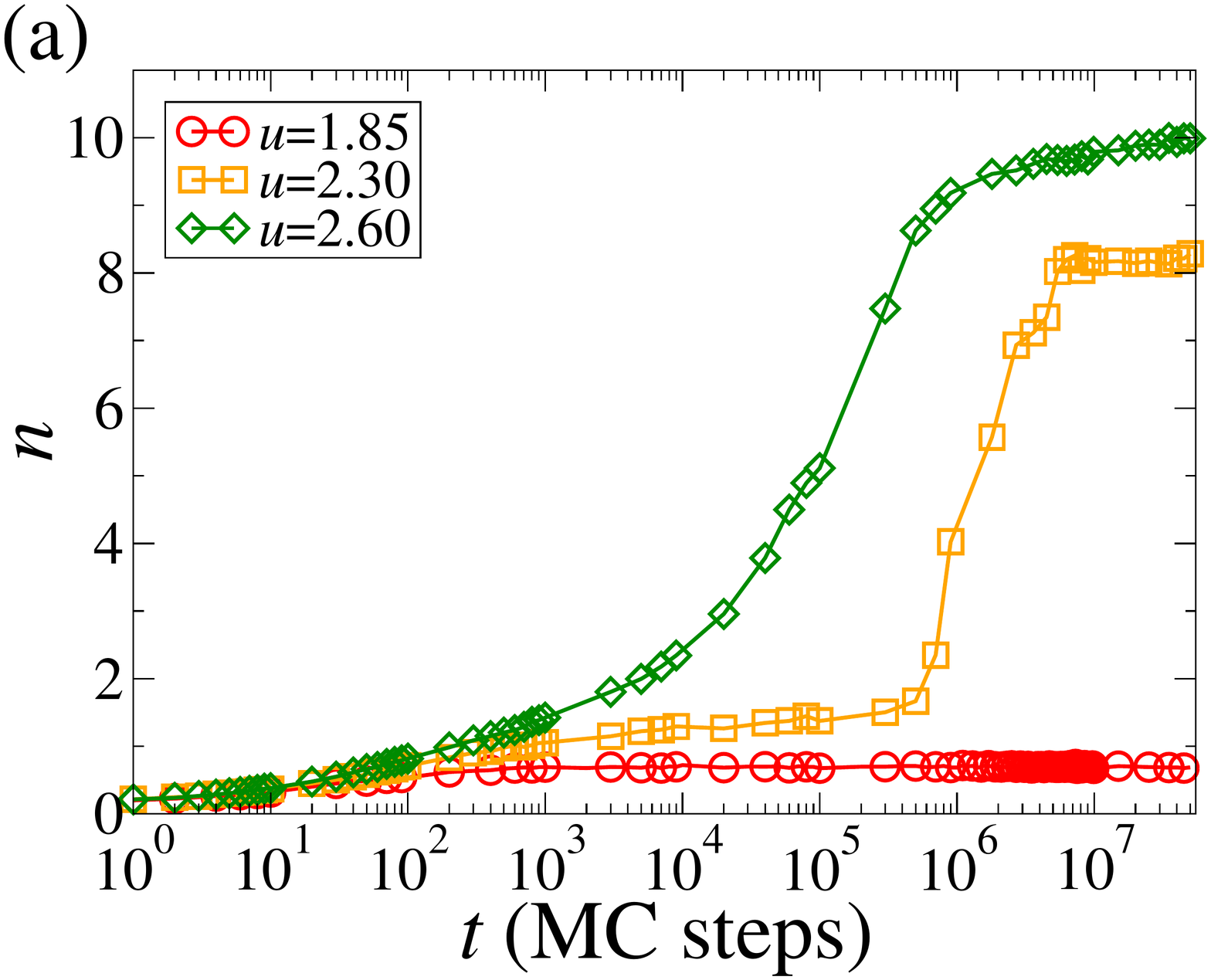}
\includegraphics[width=7.0cm]{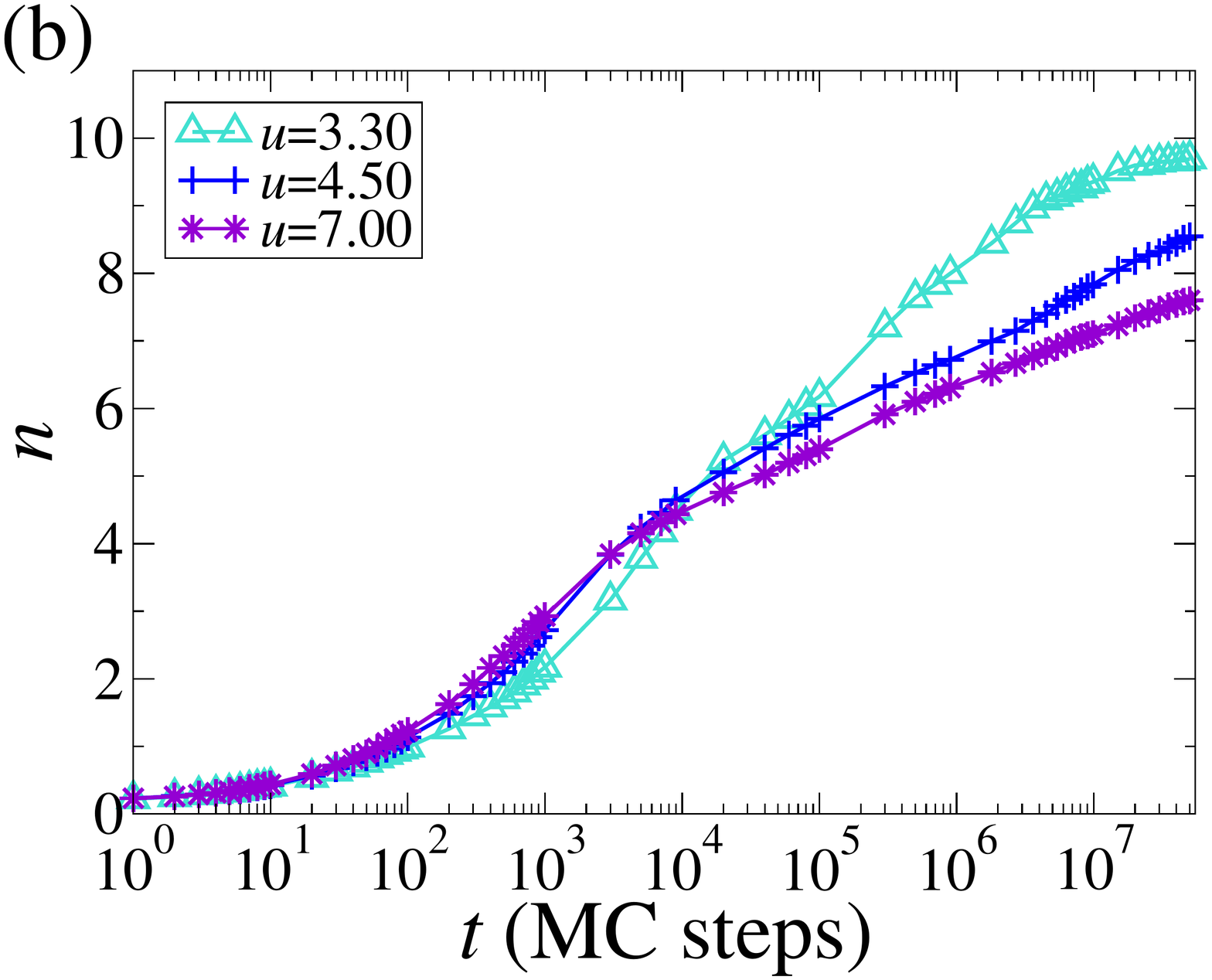}
\includegraphics[width=7.0cm]{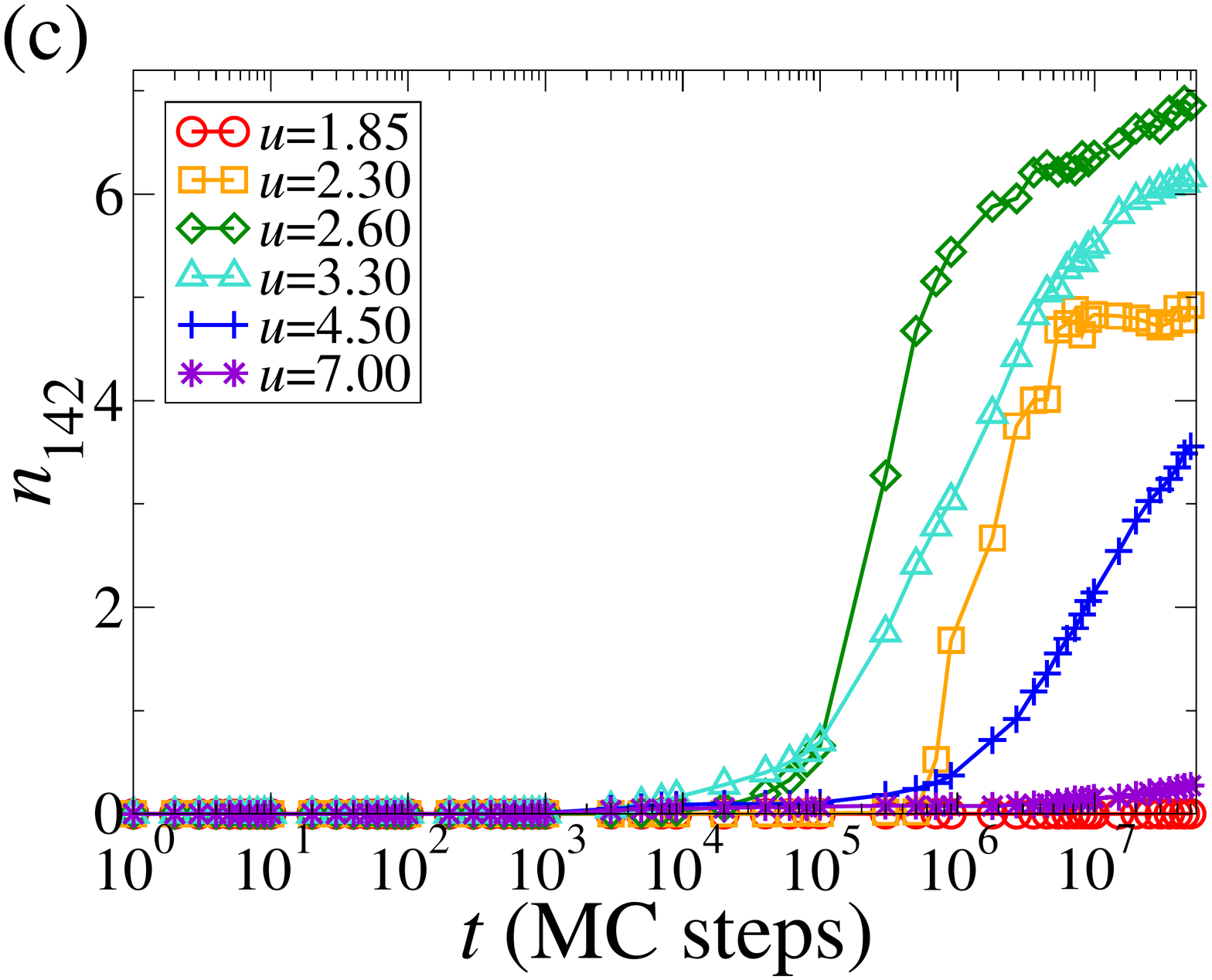}
\caption{(a,b) Average number of neighbours $n(t)$, plotted as a function of time, for different bond strengths. Each curve is averaged over eight trajectories. 
(c) The `crystallinity' of the system as a function of time for the same six bond strengths. The definition of the order
parameter $n_{142}(t)$
is discussed in the text. Each curve is an average over eight trajectories.}
\label{fig:long-time}
\end{figure}

In Fig.~\ref{fig:long-time}, we show results from dynamical simulations for various bond strengths $u$.
In the stable fluid phase (e.g., $u=1.85$), the system quickly equilibrates in a fluid state
where the average number of neighbours per particle 
\begin{equation}
n(t)=\langle n_i(t) \rangle
\end{equation} 
is relatively small.  Increasing the bond strength to $u=2.3$, the system is in the
fluid-solid phase coexistence region of the phase diagram, 
and nucleation is observed at a time $t\approx 6\times10^5$ MC steps.  
For stronger bonds, $u=2.6$ and $u=3.3$, the behaviour of $n(t)$ shows that
the nucleation barrier is small so that clusters of particles grow smoothly, starting at early times.  
For very strong bonds $u\geq4.5$,
clusters of particles grow rapidly at early times, but this growth slows down at longer times. This is the kinetic trapping regime. 

Extrapolating from the results of Liu \emph{et al.}~\cite{liu05}, we believe that the state point $u=1.85$ is in the dilute fluid phase (outside the fluid-crystal binodal) while the point $u=2.3$ is close to the metastable liquid-vapour binodal.  For simulations at $u=2.2$ (not shown), no nucleation was observed for times up to $5\times10^7$ MC steps.  This is consistent with earlier observations~\cite{frenkel97,fortini08} that nucleation is rapid in the vicinity of the metastable binodal but slow in the region between crystal-fluid and fluid-fluid binodals.

We use a common neighbour analysis (CNA)~\cite{hon87} to measure the crystallinity of the assembling system.  In particular, we
count the number of bonded pairs of particles that obey a `crystallinity criterion'.  The criterion is that pairs of particles
have exactly four mutual neighbours, and 
those mutual neighbours must share exactly two bonds.  We denote the number of bonded pairs that satisfy this criterion
by $N_{142}$, and we normalise it by defining  $n_{142}(t)=2\langle N_{142}(t)\rangle/N$. This normalisation facilitates comparison with the number of neighbours $n(t)$ plotted in Figs.~\ref{fig:long-time}(a) and (b): if all bonded pairs in the system satisfy our `crystallinity criterion' then
$n_{142}(t)=n(t)$. In the notation of Honeycutt and Andersen~\cite{hon87}, we are measuring the combined number of 1421 and 1422 environments, which are indicative of cubic crystal structures.

In Fig.~\ref{fig:long-time}(c) we show $n_{142}(t)$ for the same six indicative bond strengths considered so far. At early times $n_{142}$ is small, with a sudden increase at later times, as crystallites form in the system. Taking the data
from this figure at $t=10^6$ MC sweeps, we obtain the `yield' shown in Fig.~\ref{fig:phase-yield}. (Plotting the yield at later times shows
similar results and is discussed in Section~\ref{sec:robust}).

To summarise the results of Fig.~\ref{fig:long-time}: for $u\leq1.85$, the system does not assemble. We identify a range of good assembly between $u$=2.3 to 3.3 with the optimum around $u$=2.4 to 2.8. For energies $u$=4.5 and above we consider the system to be dominated by kinetic trapping. 

\begin{figure}
\centering
\includegraphics[width=7.5cm]{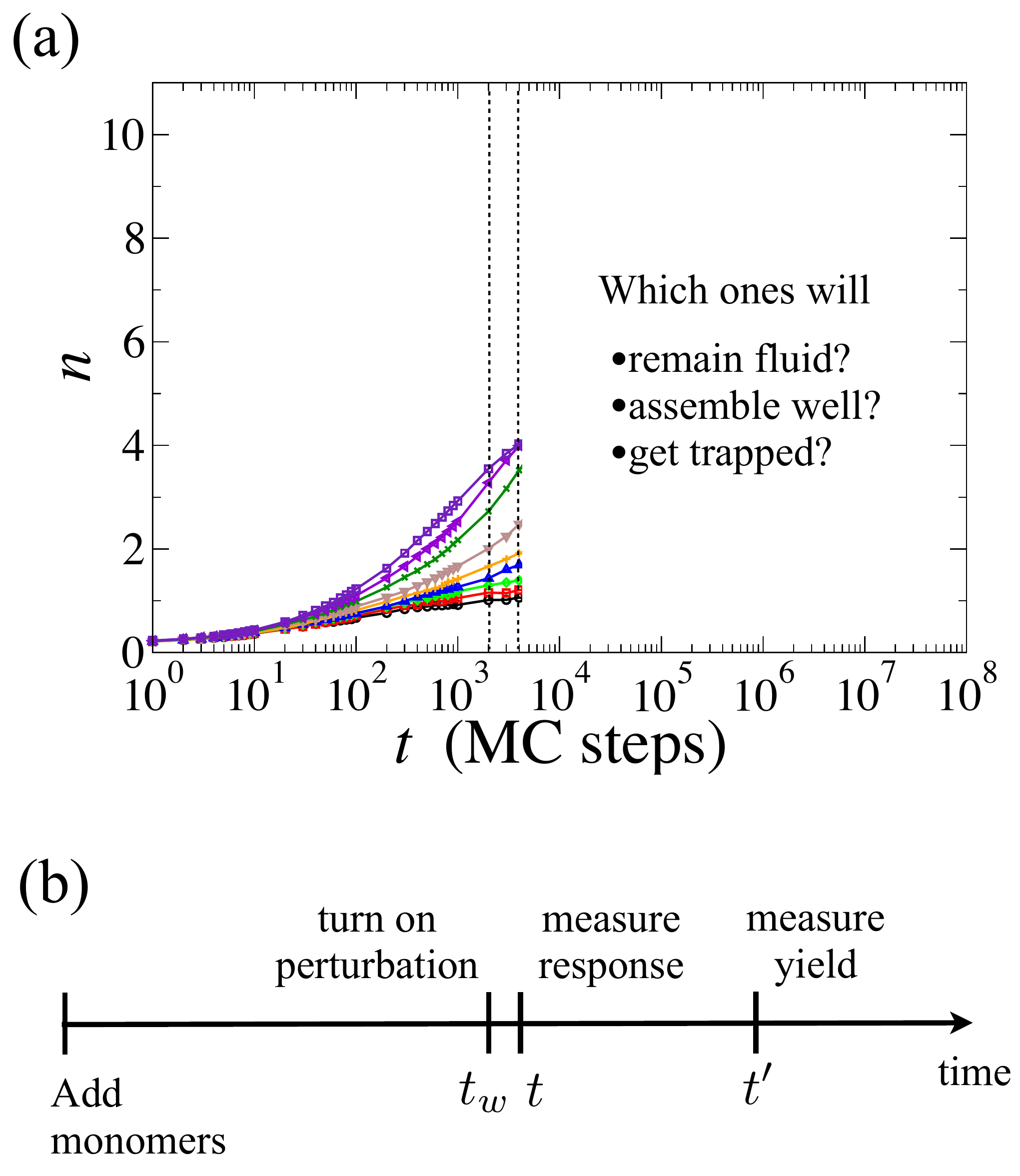}
\caption{(a) We plot $n(t)$ for various bond strengths, including the ones shown in Fig.~\ref{fig:long-time}, with the colours and symbols mixed up, and showing only $t\leq4000$ MC sweeps. From this information alone, we aim to predict which systems will remain fluid, which will assemble into crystalline structures  and which ones will get trapped. The dashed lines define the narrow window over which the dynamical measurements will be made. (b)~The simulation protocol used to measure the correlation and response functions, as described in the main text.
}
\label{fig:predict-prot}
\end{figure}

In Fig.~\ref{fig:predict-prot}(a), the number of bonds is plotted as a function of time for various temperatures, including the ones we have been focussing on up to now-- the colours and symbols of the data are now mixed. The graph has been cropped so as to show the evolution of the system only up to 4000 time steps. We pose the following question: if we are not prepared to wait for the late-time information discussed earlier in this section, can we determine the fate of the system by looking at the dynamics so early on? More specifically, which systems will remain fluid, which ones will assemble well, and which ones will get trapped? In the remainder of the paper we demonstrate that the only information needed is, in fact, in the narrow window indicated by the dashed lines between 2000-4000 MC steps.  We find that the long-term fate of the system is strongly correlated with certain early-time measurements that we will describe.

\section {Predicting assembly quality}
\label{sec:predict}

\subsection{Reversible bonding, correlation and response functions}

We aim to predict assembly quality, assuming
that all system information is available for $2000 < t < 4000$ MC sweeps. 
The hypothesis is that the dynamics of the system at early times contains enough information on whether the system will assemble and how well.  (The specific time range $2000-4000$ MC sweeps is chosen only for concreteness: 
the dependence of our results on the choice of time window is discussed below.)  A central task is to identify \emph{the propensity of the system for assembly or kinetic trapping}.
Kinetic traps arise through bonds which do not allow for long range crystalline order. Whitesides~\cite{white02} observed that the reversible formation of weak bonds allows the system to escape from kinetic traps, facilitating good assembly.  Conversely, if attractive interactions between particles are too strong, then particles tend to aggregate into disordered clusters that grow rapidly and do not anneal into crystalline structures. Our goal is to make measurements that exploit the relationship between bond-breaking and effective assembly,
in order to distinguish which systems will assemble and which will suffer from kinetic trapping.

To this end, we consider correlation and response functions that depend on the behaviour of the system between two times.  
Away from equilibrium, correlation and response measurements have been investigated 
in the theoretical study of glassy systems \cite{cuglia97,crisanti03,kurchan05}, in theories of non-equilibrium processes~\cite{baiesi09,seifert10}
as well as in self-assembling \cite{jack07} and gelating systems \cite{russo10}.  
Moreover, experimental developments have shown that measuring correlation-response functions may be a possible route for 
characterising dynamics in the lab too \cite{bonn07a, ruocco10, oukris10}. 

\begin{figure}
\centering
\includegraphics[width=7.5cm]{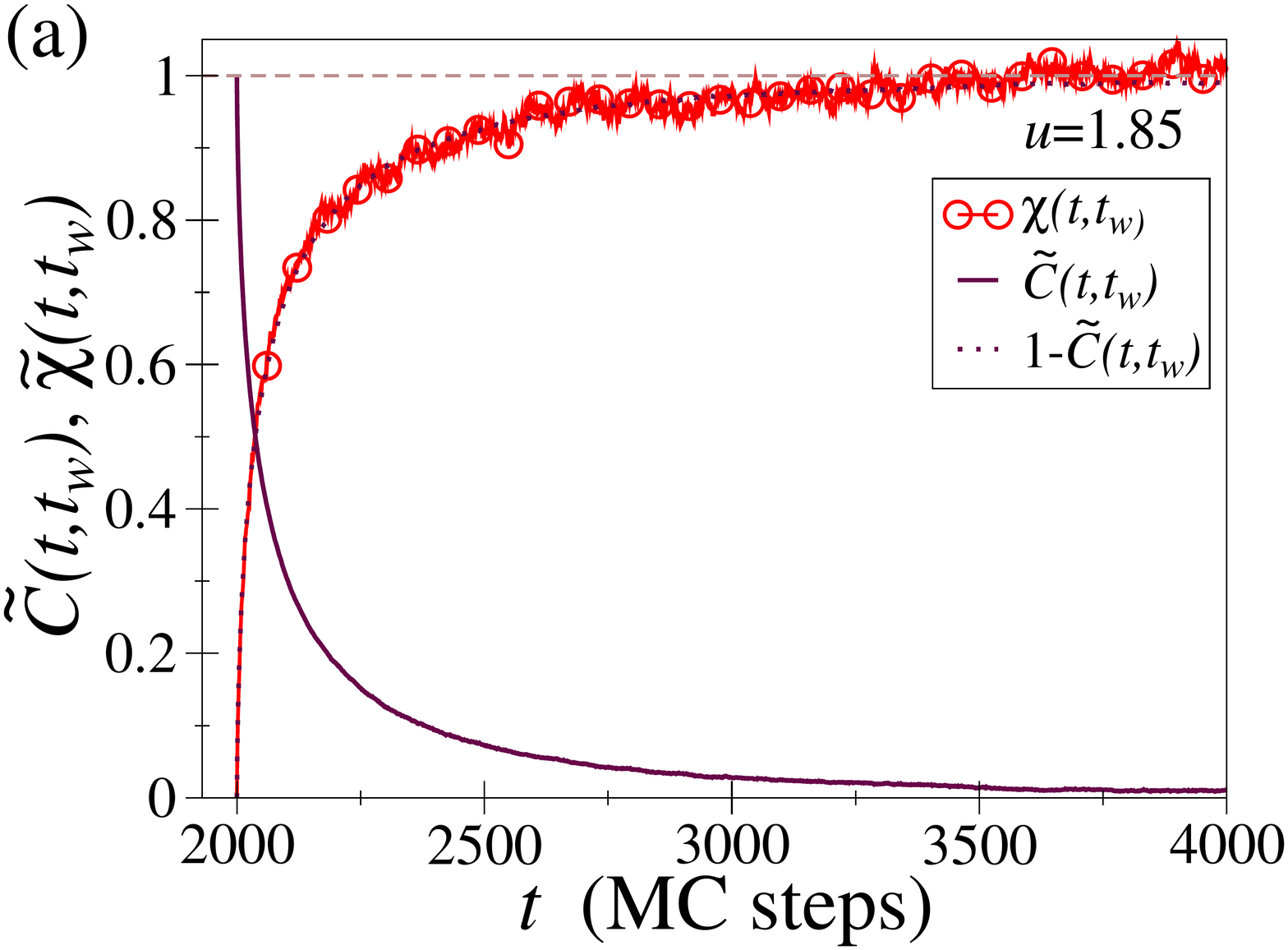}
\includegraphics[width=7.5cm]{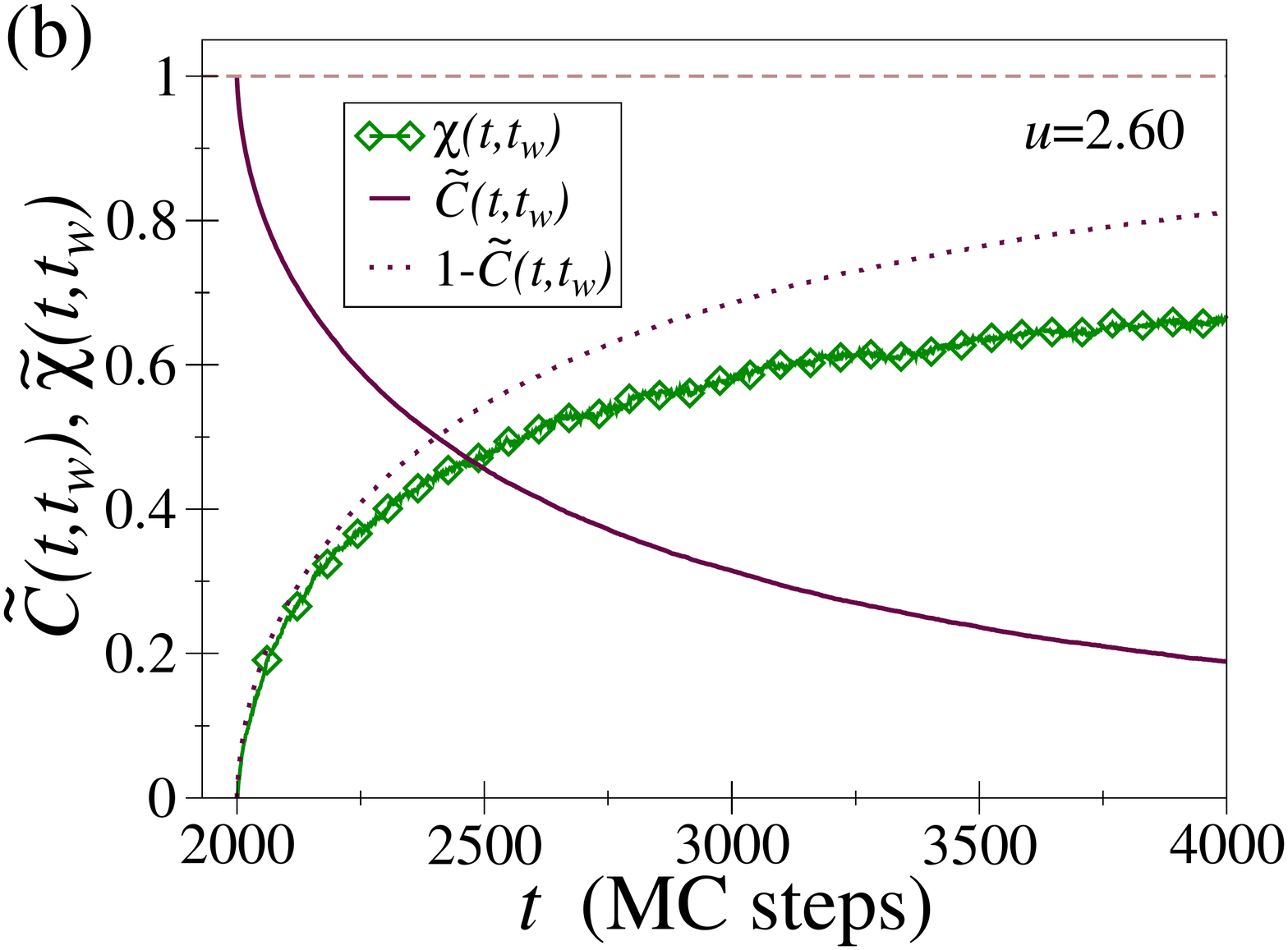}
\includegraphics[width=7.5cm]{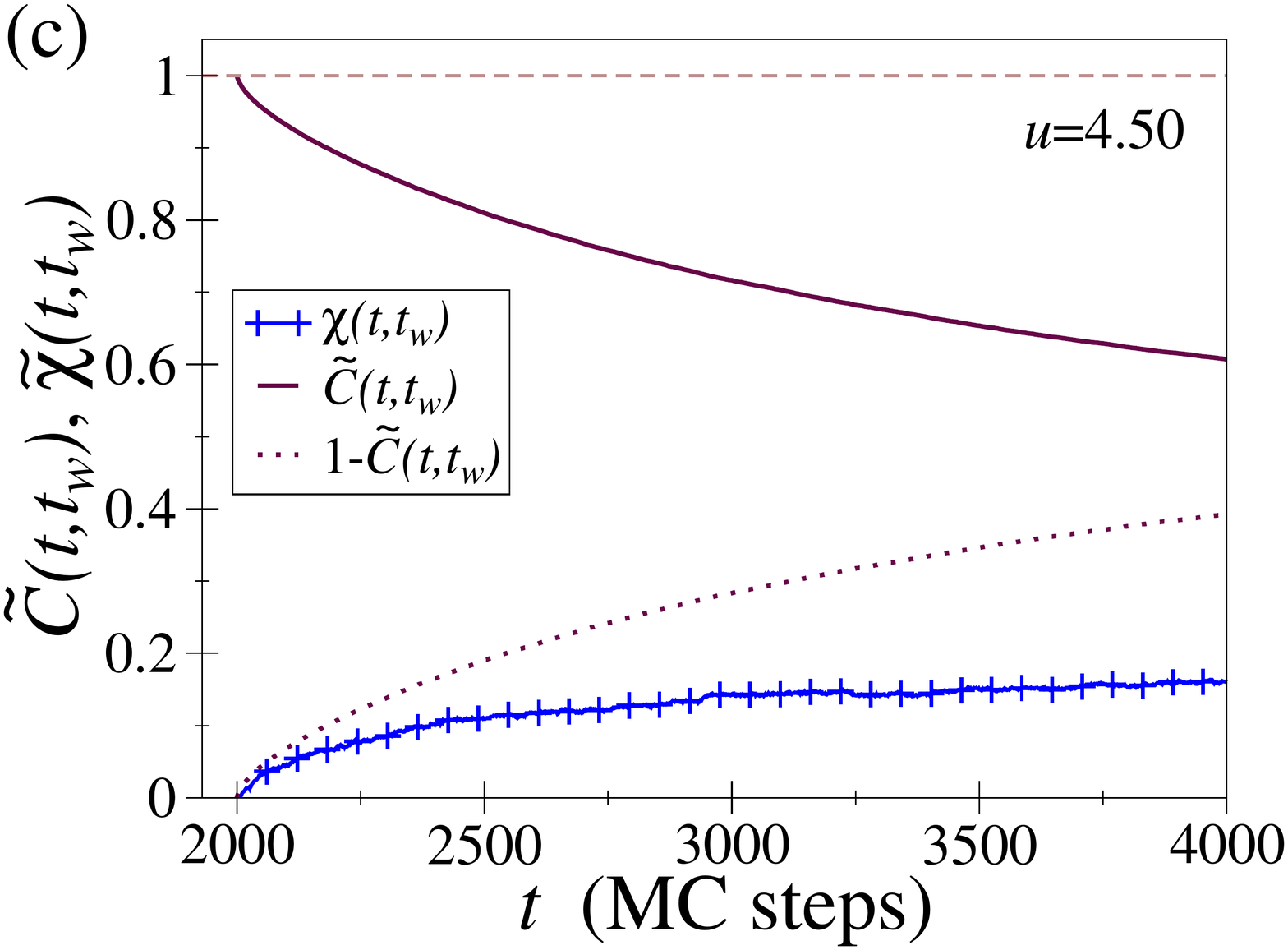}
\caption{Correlation $\tilde{C}(t,t_w)$ (solid lines) and response functions $\tilde{\chi}(t,t_w)$ (symbols) are plotted versus time $t$ for three different bond strengths. We also plot $1-\tilde{C}(t,t_w)$ (dotted line), which would be equal to the response if the system were at equilibrium. (a) Bond strength $u$=1.85: the system is equilibrated in a dilute fluid phase. The response is shown in red circles it follows the dotted line (i.e., $\tilde{\chi}(t,\tw) = 1-\tilde{C}(t,\tw)$ in accordance with FDT). (b,c) Bond strengths $u=2.6$ and $u=4.5$ respectively: both systems are far from equilibrium and the responses (symbols) differ from the plots of $1-\tilde{C}(t,\tw)$ (dotted lines).  For the weaker bonds (panel b), the
system will assemble into a crystal; for the stronger bonds (panel c), the system will become kinetically trapped.}
\label{fig:correl-resp}
\end{figure}

We define the (dimensionless) single-particle
energy autocorrelation function as
\begin{align}
C(t,t_w)& =\frac{1}{U^2} \bigl[ \langle E_i(t_w) E_i(t) \rangle - \langle E_i(t_w) \rangle \langle E_i(t) \rangle \bigr]
\nonumber \\
&=\tfrac{1}{4} \bigl[ \langle n_i(t_w) n_i(t) \rangle - \langle n_i(t_w) \rangle \langle n_i(t) \rangle \bigr]
\label{eq-corr}
\end{align}
We adopt the convention that $\tw\leq t$, consistent with Fig.~\ref{fig:predict-prot}(b).
For fixed $\tw$, the correlation function measures the extent to which the system's structure at $\tw$ is correlated
with its structure at some later time $t$.  For example, in the equilibrium fluid state the $C(t,\tw)$ decays to zero on a time
scale that reflects the lifetime of an interparticle bond.

Away from equilibrium, it is convenient to normalise this correlation function as 
\begin{equation}
\tilde{C}(t,\tw) = \frac{C(t,\tw)}{C(t,t)}
\end{equation}
The equal time correlation function $C(t,t)$ measures the variance in the number of bonds between particles, at time $t$.  If bonding
is irreversible (bonds never break), $\tilde{C}(t,\tw)$ may be interpreted as the fraction of bonds in the system at time $t$ that had
already been formed at the earlier time $\tw$.  Increasing $t$ at constant $\tw$, the system forms new bonds, and
the correlation $\tilde{C}(t,\tw)$ decays towards zero.  Thus, in contrast to the 
situation at equilibrium, the correlation function away from equilibrium does not simply measure a bond lifetime, but a combination of a bond lifetime and the rate
of bond formation.

In order to separate these effects, we also consider a response function.  The idea is that bond-making occurs whenever particles collide, so their
rates are largely independent of the bond strength.  On the other hand, bond-breaking processes require thermal activation over an energy barrier $U$
and take place with rates proportional to $\ee^{-U/\kB T}$.  On changing the strength of the forces between particles, the bond-breaking rate will change:
measuring this response gives information about the relative likelihood of bond-making and bond-breaking in the system.
The protocol for measuring this response is shown in Fig~\ref{fig:predict-prot}(b). As before, we begin with an equilibrated system of hard spheres, and introduce the interaction potential $U$ at time $t=0$. After a waiting time $\tw$ we perturb the bond strengths in the system, so that
the energy of the $i$th particle becomes $E_i=-\frac{1}{2}(U+\delta U_i)n_i$, where $n_i$ is the number of neighbours of particle $i$ (as above), while $\delta U_i$ is the perturbation applied to the $i$th particle, and the factor of $\frac{1}{2}$ ensures no double counting of bonds.  At a later
time $t> t_w$, we measure the (dimensionless) response of particle $i$ to the change in its bond strength:
\begin{equation}
\chi(t,\tw) =  \frac{\partial  \langle n_i(t) \rangle}{\partial (\delta U_i)} \frac{T}{2}.
\label{equ:resp}
\end{equation}
This response depends on $\tw$ since the perturbation is applied only between times $t$ and $\tw$.  In our simulations 
$t$ and $\tw$ vary within the window of 2000-4000 time steps as indicated by the dashed lines in Fig.~\ref{fig:predict-prot}(a). 

As shown in Appendix~\ref{app:resp}, 
$\chi(t,\tw)$ can be measured
computationally by applying perturbations of magnitude $\delta U_i$ to all particles, but with randomly chosen signs.  The response is then obtained
by comparing those particles with positive and negative values of $\delta U_i$.  We take $|\delta U_i| = 0.125U$ and have checked that for all bond strengths considered the perturbation is within the linear response regime.

In Appendix~\ref{app:fdt}, we prove the fluctuation-dissipation theorem (FDT) that links $C(t,\tw)$ and $\chi(t,\tw)$. Normalising the response as
\begin{equation}
\tilde{\chi}(t,\tw) = \frac{\chi(t,\tw)}{C(t,t)},
\end{equation}
the FDT reads
\begin{equation}
\tilde{\chi}_\mathrm{eqm}(t,t_w)=1-\tilde{C}_\mathrm{eqm}(t,t_w)
\label{equ:fdt}
\end{equation}
where we added the label `eqm' to emphasise that this relation holds only at equilibrium.

Fig.~\ref{fig:correl-resp}(a) shows correlation and response functions that are typical for a system in the dilute fluid phase at equilibrium (the bond strength is $u=1.85$). The perturbation is applied at $\tw=2000$ MC steps. The correlation function, is maximal at $t=t_w$ and then decays to zero, showing that the system eventually loses all memory of the bonds it made at time $\tw$. The response to the applied perturbation, starts to grow immediately after the perturbation is applied.   
We also plot $1-\tilde{C}(t,t_w)$ which is equal to the measured response, in accordance
with FDT, since the system is at equilibrium. 

Assembling systems, however, are far from equilibrium. Nevertheless, FDT can provide a useful comparison quantifying \emph{how far} the systems deviate from ``locally equilibrated'' states~\cite{jack07}.  From the previous section we know that the system with $u=2.6$ will assemble into a crystal, within $\sim 10^6$ MC steps. In Fig. \ref{fig:correl-resp}(b) we show correlation and response functions at times of order $4000$ MC steps, long before crystal formation. 
The decay of the correlation is slower than for $u=1.85$ and it does not reach zero within the range of times shown, indicating that the system retains some memory of the bonds it made at $\tw$.  It is also clear that the two curves $\tilde{\chi}(t, \tw)$ and 1-$\tilde{C}(t,\tw)$ no longer trace each other. 
In Fig.~\ref{fig:correl-resp}(c) we show the same plot for $u=4.5$. The correlation function decays very slowly indeed. The response is almost zero and there is a large deviation between the response $\tilde{\chi}(t, t_w)$ and 1-$\tilde{C}(t,\tw)$. A near-zero response implies minimum unbinding and therefore little annealing of disordered bonds. The bonds that were made at $\tw$ persist for long times, throughout the 2000 time-step window shown here. This indicates that the system will eventually get kinetically trapped, as was indeed shown in the previous section. 

The significance of the above observations is that we have identified a measurement which distinguishes between different far-from-equilibrium regimes of behaviour, using only information from much earlier times, at which the system bears little structural resemblance to its final product. Recalling Fig.~\ref{fig:long-time}(c), the measure of crystallinity $n_{142}$ is tiny on these time scales ($t\sim10^3$) for all the bond strengths considered.  At these early times, a structural analysis of the system has little predictive power. On the other hand, the data of Fig.~\ref{fig:correl-resp} change considerably as we pass between the equilibrated fluid phase (weak bonds), the good-assembly regime (intermediate bond strength) and the kinetically trapped (strong bond) regime.  

\subsection{Estimating fluctuation-dissipation ratios}
\label{sec:fdr}

\begin{figure}
\includegraphics[width=8.0cm]{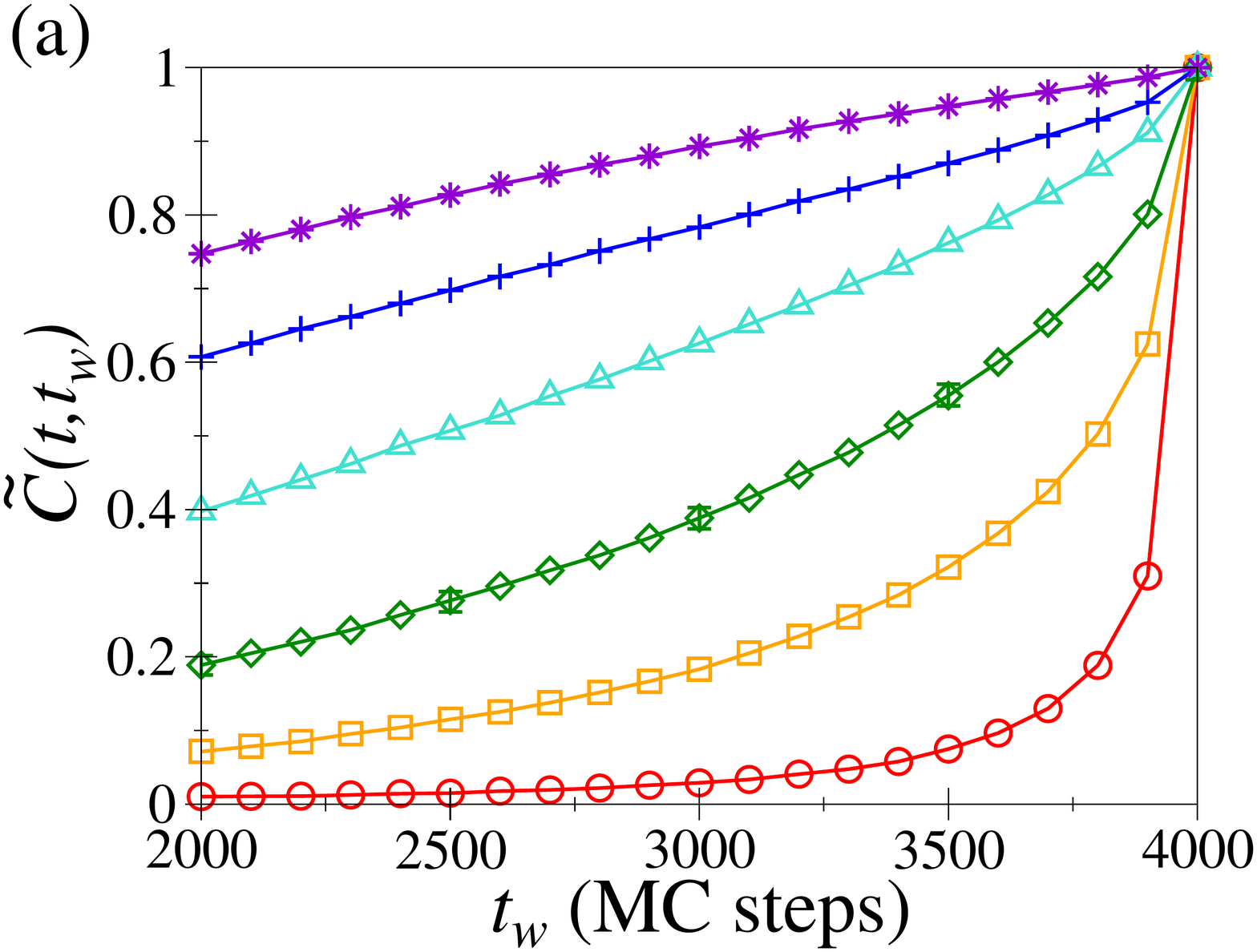}
\includegraphics[width=8.0cm]{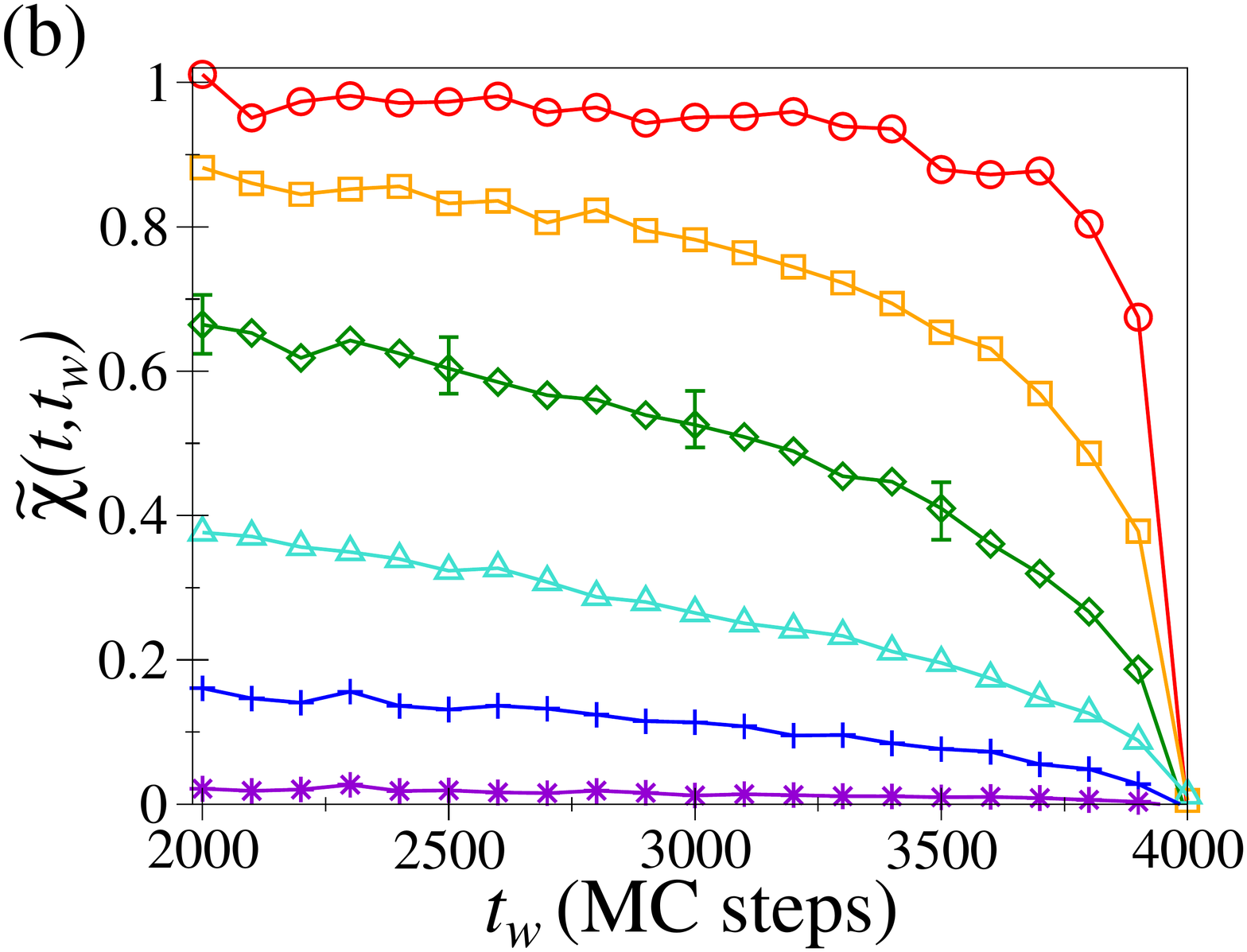}
\caption{(a) Correlation $\tilde{C}(t,\tw)$ plotted against $\tw$ for different bond strengths at fixed $t=4000$. (The symbols for the characteristic bond strengths plotted here were defined in Fig.~\ref{fig:phase-yield}(a) and are also shown in the legend of Fig.~\ref{fig:fdt}). When $t=\tw$ the correlation is maximum and decays as $\tw$ decreases (or $t-t_w$ increases). (b) Response $\tilde{\chi}(t,\tw)$ plotted against $\tw$ for different bond strengths at $t=4000$ MC steps. Correlation and response functions are estimated by averaging over many independent trajectories, with error bars shown for a few representative points in each panel.}
\label{fig:C-chi-tw}
\end{figure}

In the context of glassy model systems, 
the clearest way to analyse correlation-response data is to fix the time $t$ and vary $\tw$ \cite{sollich02,jack-fdt06,russo10}. 
Note that this is not the case in Fig.~\ref{fig:correl-resp} where correlation and response data are plotted as a function of $t$ at fixed $t_w$.
Recalling Fig.~\ref{fig:predict-prot}(b), we now fix the time 
$t$ at which the measurement is made and vary the time $\tw$ at which the perturbation is switched on.  

Results for correlation and
response functions are shown
in Fig.~\ref{fig:C-chi-tw}(a) and (b) respectively.  For $t=\tw$, the response
vanishes since the perturbation has had no time to act; as $\tw$ decreases towards zero, there is an increase in 
the time $t-\tw$ over which the perturbation acts, 
so the response grows (going from right to left in Fig.~\ref{fig:C-chi-tw}(b)).  The gradient $\partial\chi/\partial \tw$ has an interpretation as the response
of the system to an instantaneous (impulse) perturbation applied at $\tw$. 
The only difficulty when obtaining data with fixed $t$ and variable $\tw$ is that each value of $\tw$ requires a separate 
computer simulation because the time at which the perturbation is applied is different in each case.

\begin{figure}
\centering
\includegraphics[width=8.0cm]{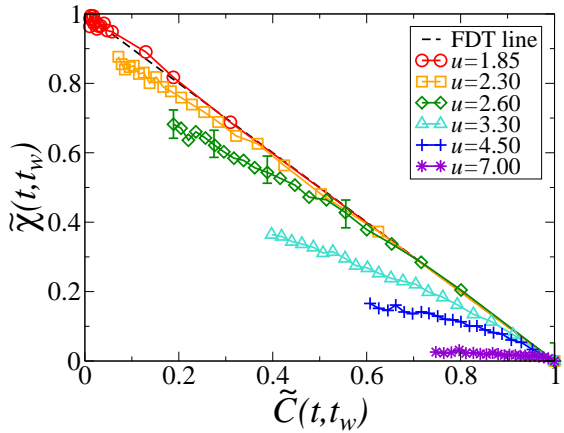}
\caption{Parametric plot of $\tilde{\chi}(t,\tw)$ versus $\tilde{C}(t,\tw)$. The data are plotted for fixed $t$=4000 MC steps, varying $\tw$ between 2000-4000 MC steps, in steps of 100. The dashed line is $\tilde{\chi}(t,t_w)$=1-$\tilde{C}(t,t_w)$, which is the FDT behaviour at equilibrium. Deviation of the data from the FDT line indicate deviations from local equilibration (see main text). 
}
\label{fig:fdt}
\end{figure}

In Fig.~\ref{fig:fdt}, we summarise the data of Fig.~\ref{fig:C-chi-tw} by making a parametric plot of the response $\tilde{\chi}(t,\tw)$ as a function of the correlation $\tilde{C}(t,\tw)$, keeping fixed $t=4000$ MC steps and varying $\tw$ and the bond strength. Such fluctuation-dissipation (FD) plots are often used in the study of glassy systems~\cite{cuglia97,sollich02,jack-fdt06,russo10} to summarise measurements of
correlation and response functions. The bottom right corner, where the correlation is maximum and the response zero, corresponds to $t=t_w$.  Following the curves from right to left (decreasing $\tilde{C}(t,\tw)$), the points indicate the behaviour as $\tw$ decreases. The dashed line is $\tilde{\chi}_{\rm{eqm}}(t,t_w)=1-\tilde{C}_{\rm{eqm}}(t,t_w)$, which is the FDT prediction for a system at equilibrium.
The data for the high temperature system lie on the FDT line, as expected since the system is equilibrated.  On crossing the binodal, the bond strength is sufficient to drive phase separation and assembly.  In this regime, the data lie on the FDT line when $t-\tw$ is small, before deviating when $\tilde C$ and $\tw$ get smaller (see for example $u=2.3, 2.6, 3.3$). The time window over which the data lie close to the FDT line decreases as the bonds become stronger. The low temperature data, $u=4.5, 7.0$, deviate from the FDT line even when $t-\tw$ is very small.

Jack \emph{et al.}~\cite{jack07},
argued that if the data are close to the FDT line when $t-\tw \approx \tau$, then the behaviour of the system is 
`locally equilibrated'\footnote{Our use of the term `local equilibration'
is similar in spirit
to an analogous condition in non-equilibrium thermodynamics~\cite{noneq-thermo-book}, but here we are referring to locality in a region of
configuration space, and not in a spatially localised region of the system. 
} on the
time scale $\tau$.
In this case, the idea is that the region of configuration space explored on these time scales 
is being explored reversibly.  That is, given a movie of the system of length $\tau$ and a similar movie where time has been reversed,
it would be difficult to discern which is which.  Clearly, if this property holds, then the likelihoods of bond-making and bond-breaking must be similar.  
On long time scales (large $t-\tw$), 
it will be apparent if a system is assembling and bond-making dominates.  But if the system is avoiding kinetic traps by exploiting the reversibility of
the bonding process in order to anneal out defects~\cite{white02} then one expects the system to be locally equilibrated over a time interval
$\tau$ that is comparable to the bond lifetime, and hence that the data in Fig.~\ref{fig:fdt} remain close to the FDT line over a significant
range of $\tilde C$.  As in Ref.~\cite{jack07}, this expectation seems to be borne out by these numerical simulations.

\subsection{Robustness of results}
\label{sec:robust}

Comparing Figs.~\ref{fig:phase-yield} and~\ref{fig:fdt}, long-time measurements of the yield are correlated with
short-time measurements of correlation and response functions.  We have in mind that the short-time measurements
might be used to predict the long-time behaviour.  However, these systems are far-from-equilibrium, and both
the short- and long-time measurements will depend on the times at which these measurements are made.  If the short-time
measurements are to be a useful predictive tool, the correlation between short-time and long-time behaviour must
be robust to variations in the time at which the measurements are made, as well as to changes in the system parameters
(for example, volume fraction and interaction range).

\begin{figure}
\centering
\includegraphics[width=8.0cm]{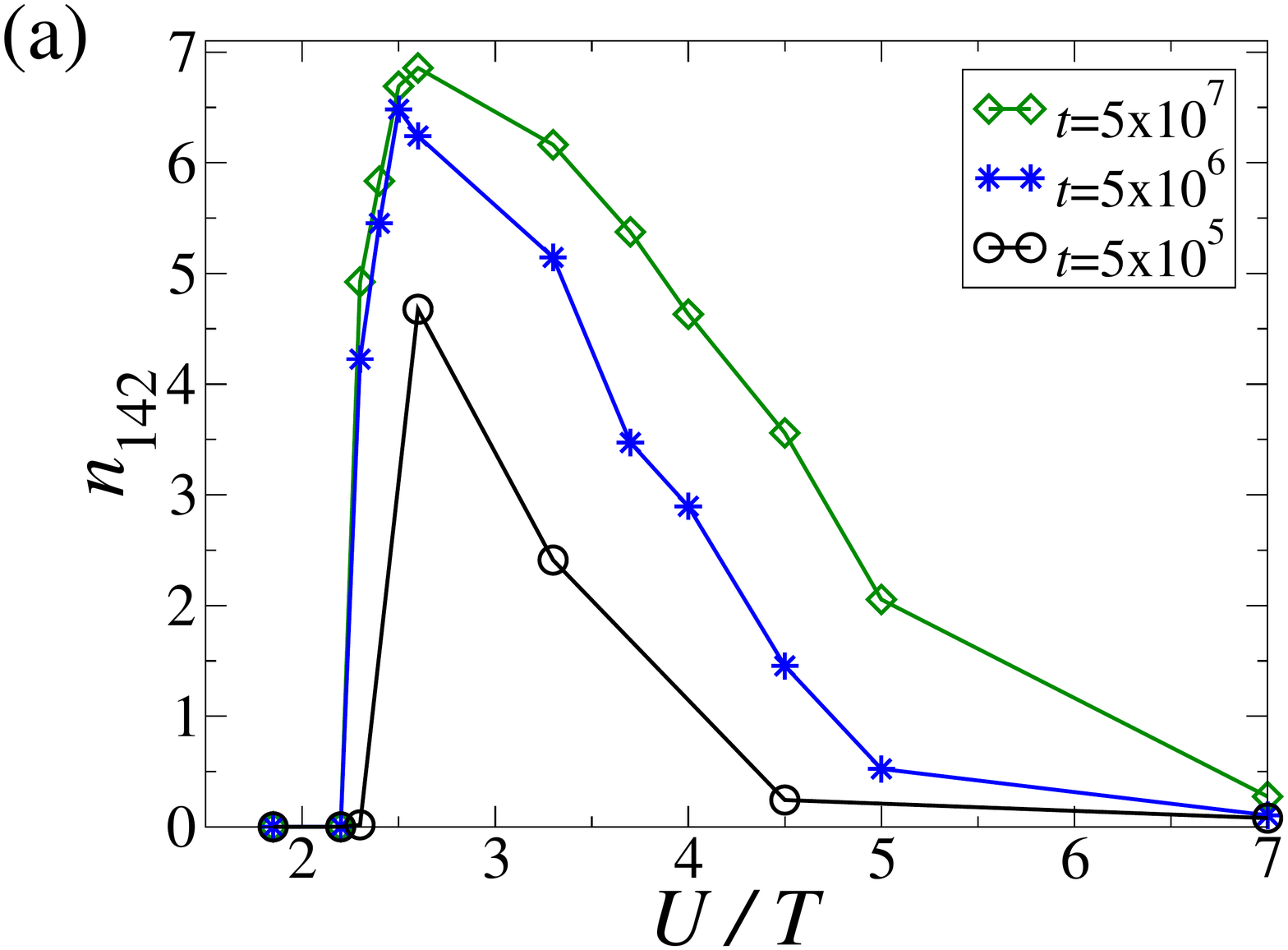}
\includegraphics[width=8.0cm]{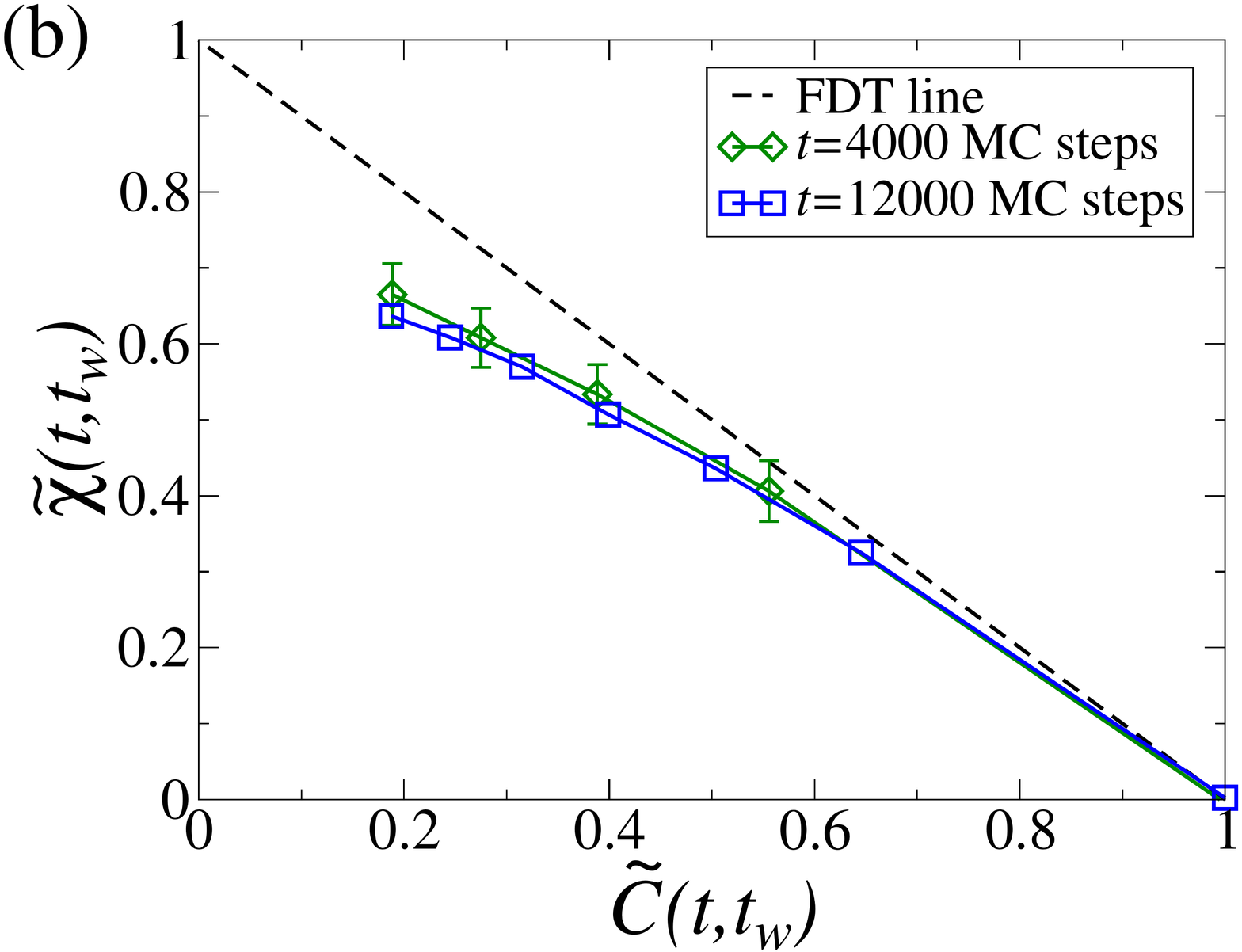}
\caption{(a)~Yield at various times, plotted as a function of the bond strength (compare Fig.~\ref{fig:phase-yield}(a)).
(b)~Parametric plot of $\tilde{\chi}(t,\tw)$ versus $\tilde{C}(t,\tw)$ for different measurement times: $t=4000$ MC steps
and $2000<\tw<4000$ MC steps as
in Fig.~\ref{fig:fdt} and $t=12000$ MC steps with $6000<\tw<12000$ MC steps. The dashed line is $\tilde{\chi}(t,t_w)$=1-$\tilde{C}(t,t_w)$, which is the FDT result at equilibrium. 
}
\label{fig:yield-t}
\end{figure}

In Fig.~\ref{fig:yield-t}, we show the effect of varying the measurement times, keeping system parameters constant.
Fig.~\ref{fig:yield-t}(a) shows that measuring the yield at different times leads to differences in the crystallinity of
the sample, as expected since the phase transformation is taking place over the whole time window considered.  
Nevertheless, a change of two orders
of magnitude in the measurement time leads to the same qualitative results, and the condition that  assembly is optimal for
$u\approx2.5$ is robust.  In Fig.~\ref{fig:yield-t}(b), we show that increasing the time at which the correlation and response
measurements are made leads to very small changes in the FD plot. (We show results for the near-optimal condition
$u=2.6$ but results are similar for other bond strengths.  Compared to Fig.~\ref{fig:fdt}, we have increased $t$ and $\tw$ by a factor
of $3$.)  This insensitivity to changes in $t$ and $\tw$
reinforces the idea that the FD plot has potential as a predictive tool,
since one arrives at the same prediction, regardless of the specific time at which the measurements are made.  

\begin{figure}
\centering
\includegraphics[width=4.25cm]{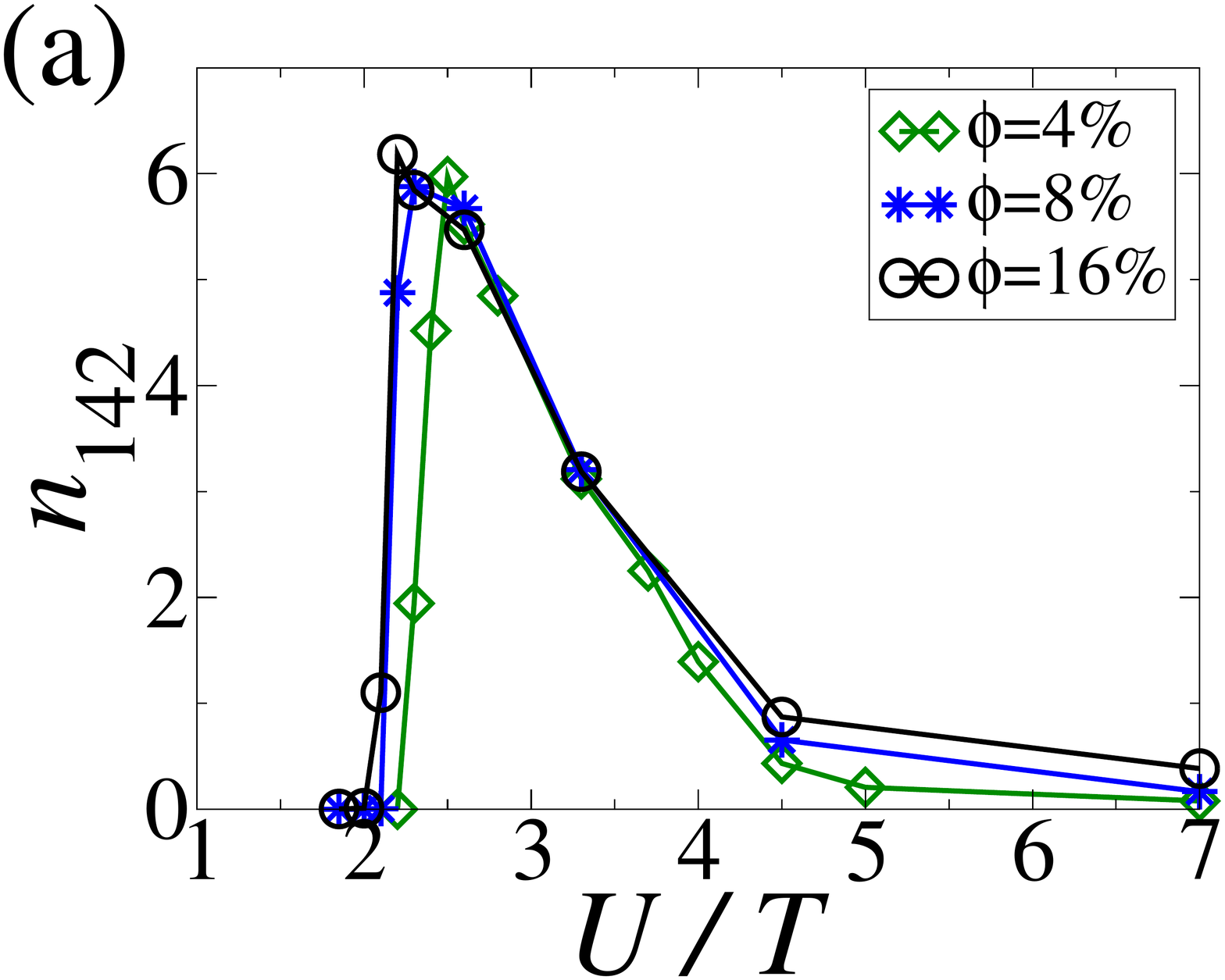}
\includegraphics[width=4.25cm]{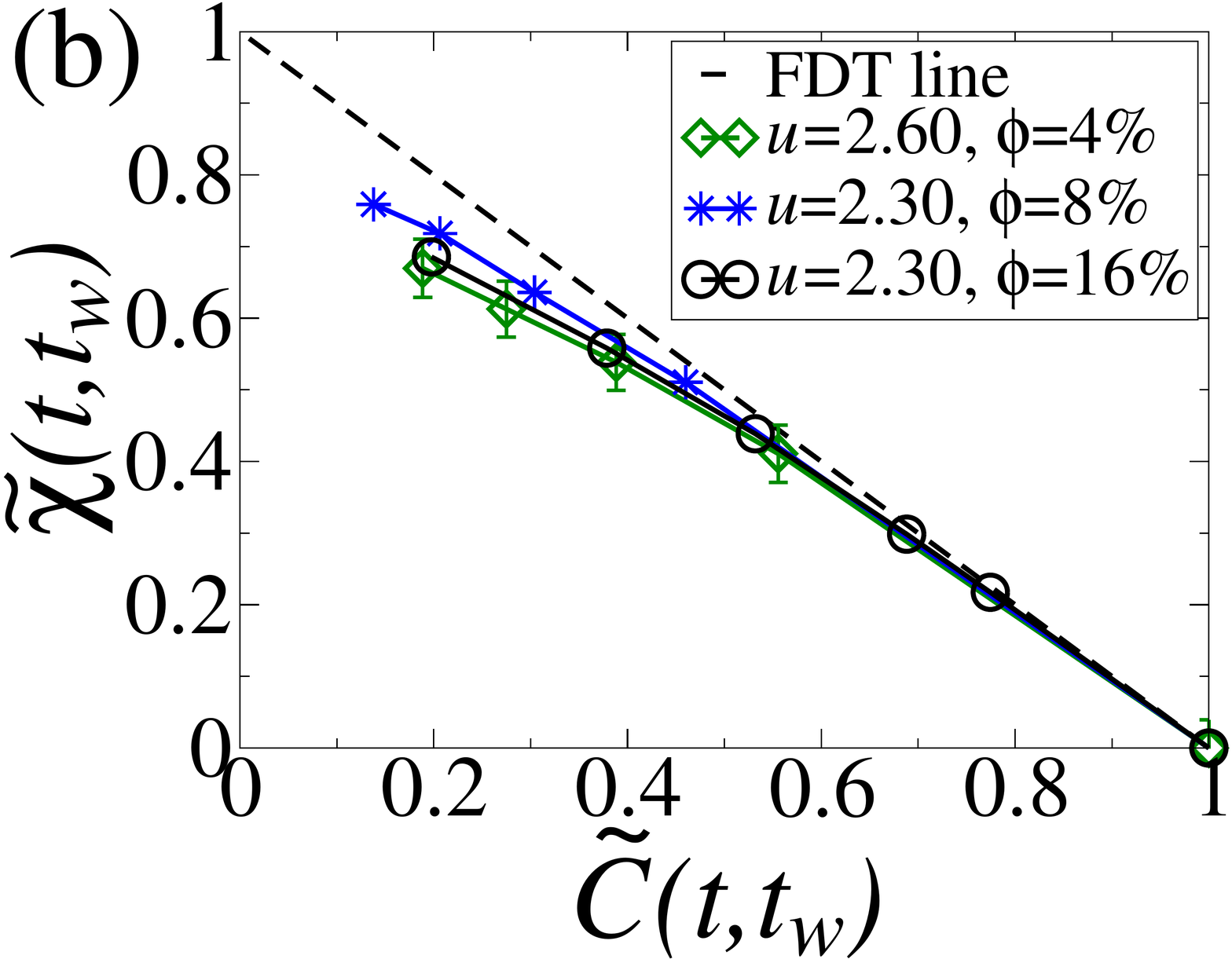}
\includegraphics[width=4.25cm]{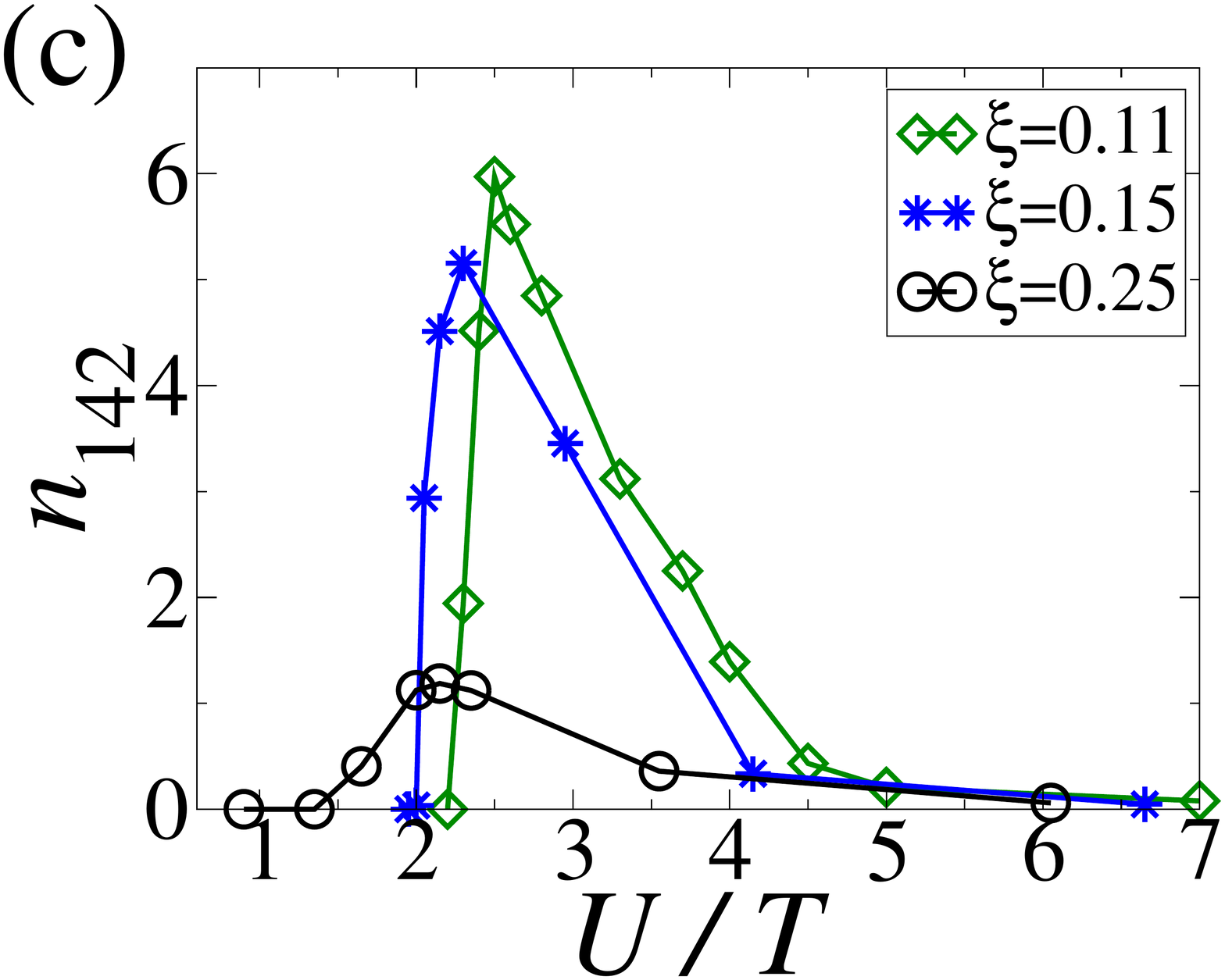}
\includegraphics[width=4.25cm]{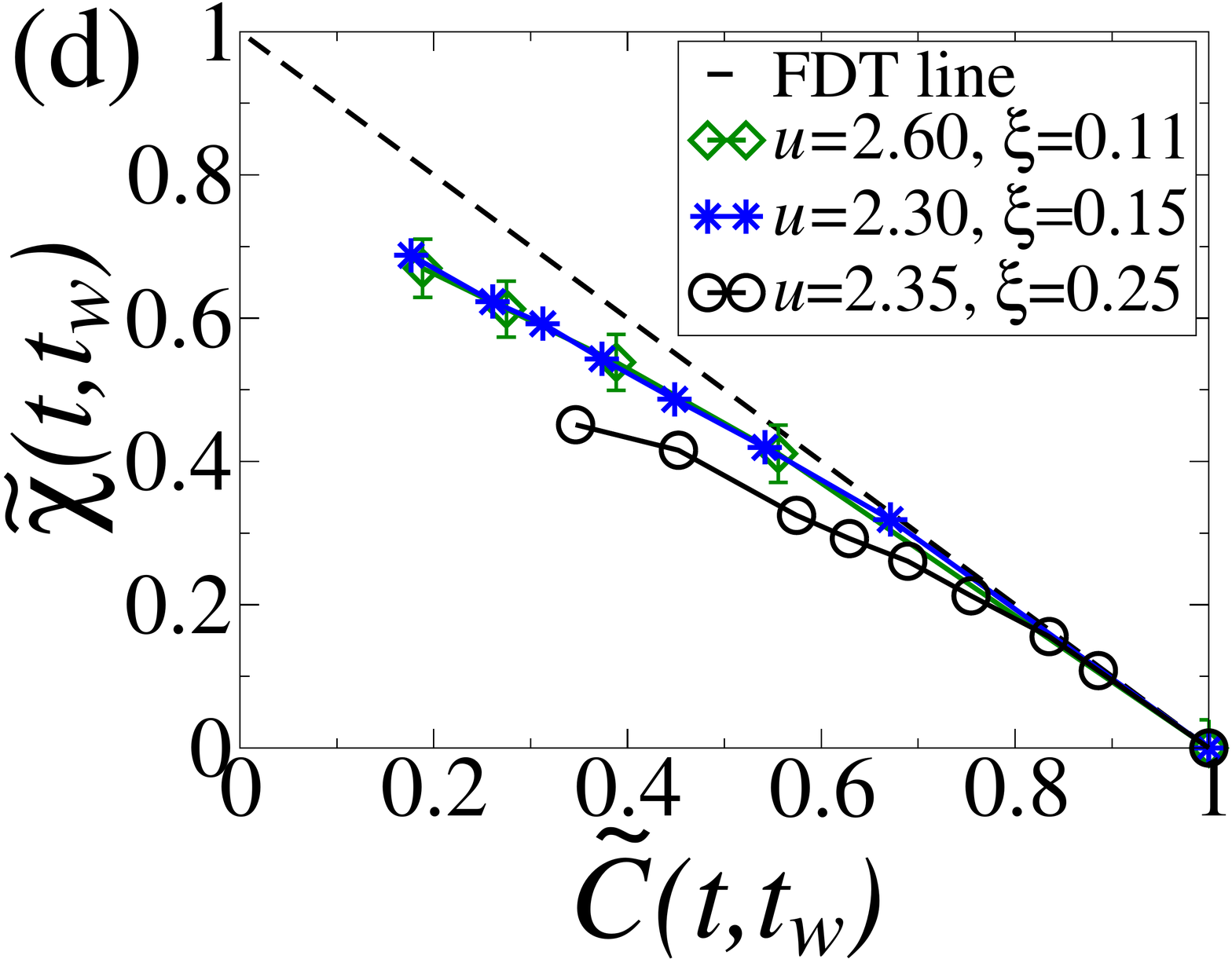}
\caption{(a)~Measurement of the yield at $t=10^6$ MC steps, as a function of the bond strength, for three different volume fractions.
(b)~FD plot, calculated at optimal assembly $u=2.6$, for two three different volume fractions.
(c)~Yield measurements for three different interaction ranges.
(d)~FD plot, calculated using the optimal value of $u$ associated with each value of $\xi$.  While the value of $u$ changes, the
FD plot for optimal $u$ is qualitatively unchanged.
}
\label{fig:yield-fdt-ksi-vol}
\end{figure}

We now discuss dependence on the volume fraction $\phi$ and the interaction range $\xi$.  
Results are summarised in Fig.~\ref{fig:yield-fdt-ksi-vol}.  On increasing the volume fraction from 4\% to 16\%, the yield (crystallinity
at long times)
 varies only slightly.  Increasing $\phi$ at constant $u$ increases the supersaturation and hence speeds up nucleation, but
this effect is rather weak in this system, for these times.  In the simulations of Fortini~\emph{et al.}~\cite{fortini08}, nucleation
was found to be very slow for values of $u$ between the crystal-fluid and fluid-fluid binodals, and they observed crystallisation
only when metastable fluid-fluid phase coexistence was also possible.  Our results are consistent with this observation,
although we were not generally able to identify long-lived metastable phase coexistence in our dynamical simulations.
In the kinetic-trapping regime, the dependence of the results on volume fraction are surprisingly weak: this is
presumably related to the thermally-activated time scales for bond-breaking, which are independent of $\phi$.
Turning to the correlation-response measurements in \ref{fig:yield-fdt-ksi-vol}(b), the dependence
on volume fraction is again weak, providing further evidence that FD plots can be used to predict conditions for effective crystallisation.

On increasing the range of the attractive interactions, the phase diagram of the system changes as the liquid phase becomes
more stable~\cite{liu05}.  For the case $\xi=0.25$, the fluid-fluid and crystal-fluid binodals are very close to each other 
(both are near to $u=1.5$ for $\phi=4\%$).
Changing $\xi$ from $0.11$ to $0.25$ has a significant effect on the yield (crystallinity) of the self-assembly process.  In particular,
for $\xi=0.25$ and a 
fixed time $t=10^6$ MC sweeps,  Fig.~\ref{fig:yield-fdt-ksi-vol}(c) shows that 
the yield rather is small for all $u$.  We find that on increasing the measurement time, the yield increases significantly, 
but the position of maximal yield remains stable at $u\approx 2$ (data not shown).  In Fig.~\ref{fig:yield-fdt-ksi-vol}(d), we show FD plots for various $\xi$, where the
value of $u$ is chosen in each case to be near near-optimal for the long-time yield.  Quantitative differences are apparent as $\xi$ is varied, but
we believe that the FD plots at optimal assembly are similar enough for predictions of effective crystallisation conditions to be made.

\section{Outlook}
\label{sec:conc}

We have considered how we might predict the evolution of a model colloidal system based on the dynamics at early times, as presented schematically in Fig.~\ref{fig:predict-prot}(a).  The data shown in Figs.~\ref{fig:fdt}, \ref{fig:yield-t} and \ref{fig:yield-fdt-ksi-vol} seem promising in this regard.
Such parametric plots can be used during the early part of the crystallisation process to distinguish between different assembly regimes, that will become structurally apparent only at later times. The current paper complements a previous study~\cite{jack07}, indicating that the method is applicable at least to patchy particles that form viral capsids, as well as to discs and spheres with short-ranged attractive interactions.  These systems have
diverse kinetically trapped states and may have no equilibrium phase transitions at all (viral capsids), or phase transitions
to both stable crystalline and metastable liquid states (as in this work).  Despite these differences,  
the correlations between parametric
plots such as Fig.~\ref{fig:fdt} and yield measurements such as Fig.~\ref{fig:phase-yield}(b) seem to be conserved between models, indicating
that the method may have broad application.

We believe that the qualitative similarity between systems is due in part to the observation of Whitesides~\cite{white02} that reversibility
is essential for effective assembly, in a wide variety of contexts. 
Essentially, for a system to assemble well into an ordered structure it has to have the ability to undo incorrect bonds, and thus avoid kinetic 
traps~\cite{white02,rap08,hagan06,jack07}. Excessively strong interactions lead to irreversible sticking and produce disordered states: this is avoided only if assembling components can adjust their positions even after they have already formed bonds. Whitesides~\cite{white02} referred to this property as reversibility and stated it as a qualitative requirement for good self-assembly or crystallisation. FD plots such as Fig. 6 constitute one route towards a \emph{quantitative} verification of this statement. 
When $t-\tw$ is small (and $\tilde C$ is large), the system is close to local equilibrium as discussed in Section~\ref{sec:fdr}: the range of $t-\tw$ over which local equilibration holds is an interval
of reversibility $\tau$.
Fig.~\ref{fig:C-chi-tw} shows that for systems that assemble effectively then $\tau$ is relatively large, while $\tau$ tends to be small for systems that eventually get arrested.  

Further, the parametric plot in Fig.~\ref{fig:fdt} provides a `dimensionless' criterion for separating these regimes: instead of 
the time $\tau$ over which the system behaves reversibly, it is natural to identify a range of $\tilde C$, which is a dimensionless variable that can be compared
directly between different systems (for example, one may compare the results of this article with those of Ref.~\cite{jack07}).  The only free parameter associated
with this comparison is the time $t$ associated with the parametric plot: we have also shown in Fig.~\ref{fig:yield-t}(a) that results for this system depend weakly on
this time, strengthening the argument that comparison of plots from different systems under different conditions can be compared fairly with each other.
In the future, we hope that the predictive information in such measurements might be useful in optimising computer simulations of assembling systems, and perhaps
even experimental crystallisation and self-assembly processes.

%% file: appendices-v3.tex
\newcommand{\deriv}[2]{\frac{\mathrm{d}#1}{\mathrm{d}#2}} 
\newcommand{\bra}[1]{\langle #1|}
\newcommand{\ket}[1]{|#1\rangle}
\newcommand{\braket}[2]{\langle #1|#2\rangle}





\numberwithin{equation}{section}
\section{Calculation of response function}
\label{app:resp}

This section explains how the response function in Eq.(\ref{equ:resp}) is calculated in simulations.  
For compactness of notation, we 
write $h_i=\delta U_i/2T$ so the response of Eq.(\ref{equ:resp}) is 
\begin{equation}
\chi(t,\tw)= 
\frac14 \frac{\partial \langle n_i(t) \rangle}{\partial h_i}
\label{eqA1-chi}
\end{equation}
where the field is applied between times $t$ and $\tw$.
We run simulations where $h_i$ is finite for all particles: we take
$h_i=|h|$ for half of the particles (chosen at random), with $h_i=-|h|$ for the other half.

%
In the presence of such a perturbation, we make a Taylor expansion of $\langle n_i \rangle_h$, where the subscript
$h$ indicates the presence of the $N$ perturbing fields $h_1,\dots,h_N$.  The result is
\begin{equation}
\langle n_i \rangle_h=\langle n_i\rangle+h_i \frac{\partial \langle n_i\rangle}{\partial h_i}+ \sum_{j\neq i} h_j\frac{\partial \langle n_i \rangle}{\partial h_j} +O(h^2)
\label{eqA1-exp}
\end{equation}
where all derivatives are evaluated at $h=0$ and we omit the dependence of $n_i$ on time $t$, for brevity.
The first-order terms in the Taylor expansion are the response of particle $i$ to its own field $h_i$, and the response of this
particle to the specific combination of other fields.  

For $j\neq i$, the response $\frac{\partial \langle n_i \rangle}{\partial h_j}$ is independent of $i$ and $j$, and scales as
$N^{-1}$ in the thermodynamic limit.  We therefore write, for $i\neq j$,
\begin{equation}
\frac{\partial \langle n_i \rangle}{\partial h_j}=\frac{c}{N}
\end{equation}
with $c=O(1)$ as $N\to\infty$ (see below).

We now decompose the sum over $j$ in (\ref{eqA1-exp}) into a contribution from those particles $j$ for which $h_j>0$, and those
with $h_j<0$.  Restricting to $j\neq i$, let $S_+$ be the set of particles $j$ for which $h_j>0$ and $S_-$ the set with $h_j<0$. 
In (\ref{eqA1-exp}), all terms in the sum are equal to either $+|h|c/N$ or $-|h|c/N$, so 
if the number of particles in $S_+$ is $N_+$ and the number in $S_-$ is $N_-$ then we have
\begin{equation}
\sum_{j\neq i} h_j\frac{\partial \langle n_i \rangle}{\partial h_j} = (N_+ - N_-) \frac{|h|c}{N}.
\end{equation}
and hence
\begin{equation}
\langle n_i \rangle_h=\langle n_i\rangle+h_i \frac{\partial \langle n_i \rangle}{\partial h_i}+|h|c \frac{N_+ - N_-}{N} +O(h^2)
\end{equation}
It is clear from this equation that we require $c=O(1)$ as $N\to\infty$ (see above)
so that the response is finite if (for example) a positive
field is applied to exactly half of the particles and the other particles are unchanged ($N_+=\frac{N}{2}$ and $N_-=0$).

In the case where half of the particles receive a perturbation of $+|h|$ and half receive $-|h|$,
 the value of $\langle n_i \rangle_h$ depends on $i$ only through the sign of $h_i$.  For particles
with $h_i>0$, we write their average number of neighbours by $\langle n_i \rangle_+$ while for
particles with $h_i<0$ we write $\langle n_i \rangle_-$.  These quantities are readily calculated in a simulation,
by evaluating the number of bonds for each particle and averaging separately over those with $h_i>0$ and those
with $h_i<0$.

In total, $\frac{N}2$ particles have $h_i>0$ and $\frac{N}2$ have $h_i<0$.
However, the sets $S_+$ and $S_-$ both exclude particle $i$, so if $h_i>0$ then
$N_+=\frac{N}{2} - 1$ while $N_-=\frac{N}{2}$.   Hence
\begin{equation}
\langle n_i \rangle_+=\langle n_i\rangle+|h| \frac{\partial \langle n_i \rangle}{\partial h_i}-|h|\frac{c}{N} + O(h^2)
\label{eqA1-plus1}
\end{equation}
Similarly if $h_i<0$, then $N_+=\frac{N}{2}$ while $N_-=\frac{N}{2}-1$, so that 
\begin{equation}
\langle n_i \rangle_-=\langle n_i\rangle-|h| \frac{\partial \langle n_i \rangle}{\partial h_i}+|h|\frac{c}{N} + O(h^2)
\label{eqA1-minus1}
\end{equation}

It is therefore clear that the average particle energy in the absence of the field may be estimated in
the perturbed system by
\begin{equation}
\langle n_i\rangle=\frac{\langle n_i \rangle_++\langle n_i \rangle_-}{2} + O(h^2)
\end{equation}
while the single particle response function can be estimated as
\begin{equation}
\frac{\partial \langle n_i \rangle}{\partial h_i} = \frac{\langle n_i \rangle_+-\langle n_i \rangle_-}{2|h|}+O(|h|)+O(1/N) 
\end{equation}
That is, the response $\chi(t,\tw)$ in Eqs.(\ref{equ:resp}) and (\ref{eqA1-chi}) can be estimated as 
$\chi(t,\tw)=\frac{T}{4|\delta U|}(\langle n_i \rangle_+-\langle n_i \rangle_-)$ by measuring the difference in the number of neighbours
of particles for which the perturbing field $\delta U_i$ has positive or negative sign.  This is the method used to calculate
$\chi(t,\tw)$ in this article.

\section{Fluctuation-dissipation theorem}
\label{app:fdt}

\newcommand{\cci}{I}
\newcommand{\ccj}{J}
\newcommand{\ccf}{F}

In this section, we prove Eq.(\ref{equ:fdt}) of the main text.  As in the previous section, we write $h_i=\delta U_i/2T$.
The
response of Eq.(\ref{equ:resp}) may be written in terms of probabilities by using (\ref{eqA1-chi}) together with
\begin{equation}
\frac{\partial \langle n_i \rangle}{\partial h_i}=
\frac{\partial}{\partial h_i} \left(\sum_{\cci,\ccf} \rho(\cci)G^h_{t-\tw}(\cci\rightarrow \ccf)n_i(\ccf)\right),
\label{equ:prob}
\end{equation}
where $\cci$ is the configuration of the system at time $\tw$ and $\ccf$ is the configuration at time $t$: the sum
runs over all possible configurations $I$ and $F$. Also, $n_i(\ccf)$ 
is the number of neighbours of particle $i$ in configuration $\ccf$; the initial distribution
$\rho(\cci)$ is the probability of being in configuration $\cci$ at time $\tw$; and
the propagator $G^{h}_{t-\tw}(\cci\rightarrow \ccf)$
is the probability of being in configuration $\ccf$ at time $t$, given that the system was in configuration $\cci$ at time $\tw$.
The label $h$ on the propagator indicates that it depends on the applied fields $h_i$, while $\rho(\cci)$ and $n_i(\ccf)$
do not.

Since we are concerned with Eq.(\ref{equ:fdt}), we restrict to the case of equilibrium response functions, for which
the system is equilibrated with $h=0$ at time $\tw$:
\begin{equation}
\rho(\cci) = \rho_{\mathrm{eqm}}^{h=0}(\cci)=\frac{\ee^{uB(\cci)}}{Z},
\end{equation}
where $Z=\sum_I \ee^{uB(\cci)}$ and $B(\cci)=\frac12 \sum_i n_i(\cci)$ is the total number of bonds in configuration $\cci$.
The energy of configuration $\cci$ in the unperturbed system is $-U B(\cci)$ while in the perturbed
system it is $-U B(\cci) - T \sum_i h_i n_i(\cci)$. Hence we define
\begin{equation}
\rho_{\mathrm{eqm}}^{h}(\cci)=\frac{1}{Z_h}\ee^{uB(\cci)+ \sum_i h_i n_i(\cci)},
\label{equ:rhoh}
\end{equation}
with $Z_h=\sum_\cci \ee^{uB(\cci)+\sum_i h_i n_i(\cci)}$ the partition function.

To prove Eq.(\ref{equ:fdt}) we make use of detailed balance.  For a single Monte Carlo step in the presence of the
perturbation $h$, let the probability
of arriving in configuration $\ccj$ from an initial configuration $\cci$ be $P^h(\cci\to\ccj)$.
Detailed balance states that
\begin{equation}
\rho_{\mathrm{eqm}}^h(\cci)P^h(\cci\to \ccj)=\rho_{\mathrm{eqm}}^h(\ccj)P^h(\ccj\rightarrow \cci),
\end{equation}
which ensures that the system converges to the equilibrium distribution in the limit of long times.
A similar relation follows for the propagator:
\begin{equation}
\rho_{\mathrm{eqm}}^h(\cci)G^h_{t-\tw}(\cci\to \ccf)=\rho_{\mathrm{eqm}}^h(\ccf)G^h_{t-\tw}(\ccf\rightarrow \cci).
\end{equation}

In (\ref{equ:prob}), the only $h$-dependence comes through $G^h$ so we seek an expression for $\frac{\partial}{\partial h_i}G^h$.
We use detailed balance together with (\ref{equ:rhoh}) to write
\begin{multline}
G^h_{t-\tw}(\cci\to\ccf)
=G^h_{t-\tw}(\ccf\to\cci) \times \\ \mathrm{e}^{uB(\ccf)-uB(\cci)+\sum_i[h_in_i(\ccf)-h_in_i(\cci)]}.
\label{eq-g}
\end{multline}
Taking a derivative with respect to $h_i$ and evaluating it at $h=0$, we arrive at
\begin{multline}
\frac{\partial}{\partial h_i}G^h_{t-\tw}(\cci\to\ccf)
= \mathrm{e}^{uB(\ccf)-uB(\cci)} \times \\  \left[ (n_i(\ccf)-n_i(\cci))  G^{h=0}_{t-\tw}(\ccf\to\cci) + \frac{\partial}{\partial h_i} G^h_{t-\tw}(\ccf\to\cci)
\right]. \\
\label{equ:Gh}
\end{multline}

Now, detailed balance implies that $\mathrm{e}^{uB(\ccf)-uB(\cci)} G^{h=0}_{t-\tw}(\ccf\to\cci) = G^{h=0}_{t-\tw}(\cci\to\ccf)$, and we
also have $\rho^{h=0}_\mathrm{eqm}(\ccf) \mathrm{e}^{uB(\cci)-uB(\ccf)} = \rho^{h=0}_\mathrm{eqm}(\cci)$.  Combining these
results with (\ref{equ:prob}) and (\ref{equ:Gh}) yields
\begin{align}
\frac{\partial \langle n_i \rangle}{\partial h_i}= & \sum_{\cci,\ccf} 
  \rho_\mathrm{eqm}^{h=0}(\cci)   G^{h=0}_{t-\tw}(\cci\to\ccf) 
   [n_i(\ccf)-n_i(\cci)]n_i(\ccf) 
\nonumber \\
   & + \sum_{\cci,\ccf} \rho_\mathrm{eqm}^{h=0}(\ccf) \frac{\partial}{\partial h_i} G^{h}_{t-\tw}(\ccf\to\cci)
   n_i(\ccf),
 \label{equ:prob2}  
\end{align}
where we recognise the first term on the right hand side as the correlation function $\langle n_i(t)[n_i(t)-n_i(\tw)]  \rangle$, evaluated
at equilibrium.  The second term on the right hand side is zero since $\sum_\cci G^{h}_{t-\tw}(\ccf\to\cci)=1$: this
follows from the definition of $G_{t-\tw}(\ccf\to\cci)$ as the probability of being in state $\cci$ at time $t$
since these probabilities must sum to unity, regardless
of $F$, $h$ and $t-\tw$.

Hence, at equilibrium, (\ref{equ:prob2}) reduces to
\begin{equation}
\frac{\partial \langle n_i \rangle}{\partial h_i} = \langle n_i(t) [n_i(t)-n_i(\tw)]  \rangle .
\label{equ:nearly-fdt}
\end{equation}

To recover Eq.(\ref{equ:fdt}) of the main text, we note from (\ref{eq-corr}) that 
\begin{equation}C(t,\tw) = \tfrac14[ \langle n_i(t) n_i(\tw) \rangle - \langle n_i(t) \rangle\langle n_i(\tw) \rangle]\end{equation} so that
the right hand side of (\ref{equ:nearly-fdt}) is $4[C(t,t) - C(t,\tw)] + \langle n_i(t) \rangle \langle n_i(t) - n_i(\tw) \rangle$.
At equilibrium, time-translational invariance implies $\langle n_i(t) - n_i(\tw) \rangle=0$, and using (\ref{eqA1-chi}) with (\ref{equ:nearly-fdt}) one arrives at the fluctuation-dissipation theorem
\begin{equation}
\chi_\mathrm{eqm}(t,\tw) = C_\mathrm{eqm}(t,t) - C_\mathrm{eqm}(t,\tw).
\end{equation}
Finally, dividing both sides by $C_\mathrm{eqm}(t,t)$ yields the normalised version of the FDT, which is given in Eq.(\ref{equ:fdt}).


%% file: aps-9.bbl
\begin{thebibliography}{50}%
\makeatletter
\providecommand \@ifxundefined [1]{%
 \@ifx{#1\undefined}
}%
\providecommand \@ifnum [1]{%
 \ifnum #1\expandafter \@firstoftwo
 \else \expandafter \@secondoftwo
 \fi
}%
\providecommand \@ifx [1]{%
 \ifx #1\expandafter \@firstoftwo
 \else \expandafter \@secondoftwo
 \fi
}%
\providecommand \natexlab [1]{#1}%
\providecommand \enquote  [1]{``#1''}%
\providecommand \bibnamefont  [1]{#1}%
\providecommand \bibfnamefont [1]{#1}%
\providecommand \citenamefont [1]{#1}%
\providecommand \href@noop [0]{\@secondoftwo}%
\providecommand \href [0]{\begingroup \@sanitize@url \@href}%
\providecommand \@href[1]{\@@startlink{#1}\@@href}%
\providecommand \@@href[1]{\endgroup#1\@@endlink}%
\providecommand \@sanitize@url [0]{\catcode `\\12\catcode `\$12\catcode
  `\&12\catcode `\#12\catcode `\^12\catcode `\_12\catcode `\%12\relax}%
\providecommand \@@startlink[1]{}%
\providecommand \@@endlink[0]{}%
\providecommand \url  [0]{\begingroup\@sanitize@url \@url }%
\providecommand \@url [1]{\endgroup\@href {#1}{\urlprefix }}%
\providecommand \urlprefix  [0]{URL }%
\providecommand \Eprint [0]{\href }%
\providecommand \doibase [0]{http://dx.doi.org/}%
\providecommand \selectlanguage [0]{\@gobble}%
\providecommand \bibinfo  [0]{\@secondoftwo}%
\providecommand \bibfield  [0]{\@secondoftwo}%
\providecommand \translation [1]{[#1]}%
\providecommand \BibitemOpen [0]{}%
\providecommand \bibitemStop [0]{}%
\providecommand \bibitemNoStop [0]{.\EOS\space}%
\providecommand \EOS [0]{\spacefactor3000\relax}%
\providecommand \BibitemShut  [1]{\csname bibitem#1\endcsname}%
\let\auto@bib@innerbib\@empty
\bibitem [{\citenamefont {Langer}\ and\ \citenamefont
  {Tirrel}(2004)}]{langer04}%
  \BibitemOpen
  \bibfield  {author} {\bibinfo {author} {\bibfnamefont {R.}~\bibnamefont
  {Langer}}\ and\ \bibinfo {author} {\bibfnamefont {D.~A.}\ \bibnamefont
  {Tirrel}},\ }\href@noop {} {\bibfield  {journal} {\bibinfo  {journal}
  {{N}ature}\ }\textbf {\bibinfo {volume} {428}},\ \bibinfo {pages} {487}
  (\bibinfo {year} {2004})}\BibitemShut {NoStop}%
\bibitem [{\citenamefont {Stupp}(2010)}]{stupp10}%
  \BibitemOpen
  \bibfield  {author} {\bibinfo {author} {\bibfnamefont {S.~I.}\ \bibnamefont
  {Stupp}},\ }\href@noop {} {\bibfield  {journal} {\bibinfo  {journal} {{N}ano
  {L}ett.}\ }\textbf {\bibinfo {volume} {10}},\ \bibinfo {pages} {4783}
  (\bibinfo {year} {2010})}\BibitemShut {NoStop}%
\bibitem [{\citenamefont {Love}\ \emph {et~al.}(2005)\citenamefont {Love},
  \citenamefont {Estroff}, \citenamefont {Kriebel}, \citenamefont {Nuzzo},\
  and\ \citenamefont {Whitesides}}]{love05}%
  \BibitemOpen
  \bibfield  {author} {\bibinfo {author} {\bibfnamefont {J.~C.}\ \bibnamefont
  {Love}}, \bibinfo {author} {\bibfnamefont {L.~A.}\ \bibnamefont {Estroff}},
  \bibinfo {author} {\bibfnamefont {J.~K.}\ \bibnamefont {Kriebel}}, \bibinfo
  {author} {\bibfnamefont {R.~G.}\ \bibnamefont {Nuzzo}}, \ and\ \bibinfo
  {author} {\bibfnamefont {G.~M.}\ \bibnamefont {Whitesides}},\ }\href@noop {}
  {\bibfield  {journal} {\bibinfo  {journal} {{C}hem. {R}ev.}\ }\textbf
  {\bibinfo {volume} {105}},\ \bibinfo {pages} {1103} (\bibinfo {year}
  {2005})}\BibitemShut {NoStop}%
\bibitem [{\citenamefont {Mastrangeli}\ \emph {et~al.}(2009)\citenamefont
  {Mastrangeli}, \citenamefont {Abbasi}, \citenamefont {Varel}, \citenamefont
  {Hoof}, \citenamefont {Celis},\ and\ \citenamefont {Bohringer}}]{mastra09}%
  \BibitemOpen
  \bibfield  {author} {\bibinfo {author} {\bibfnamefont {M.}~\bibnamefont
  {Mastrangeli}}, \bibinfo {author} {\bibfnamefont {S.}~\bibnamefont {Abbasi}},
  \bibinfo {author} {\bibfnamefont {C.}~\bibnamefont {Varel}}, \bibinfo
  {author} {\bibfnamefont {C.~V.}\ \bibnamefont {Hoof}}, \bibinfo {author}
  {\bibfnamefont {J.-P.}\ \bibnamefont {Celis}}, \ and\ \bibinfo {author}
  {\bibfnamefont {K.~F.}\ \bibnamefont {Bohringer}},\ }\href@noop {} {\bibfield
   {journal} {\bibinfo  {journal} {{J}ournal of {M}icromechanics {A}nd
  {M}icroengineering}\ }\textbf {\bibinfo {volume} {1}},\ \bibinfo {pages}
  {083001} (\bibinfo {year} {2009})}\BibitemShut {NoStop}%
\bibitem [{\citenamefont {Whitesides}\ and\ \citenamefont
  {Lipomi}(2009)}]{white09}%
  \BibitemOpen
  \bibfield  {author} {\bibinfo {author} {\bibfnamefont {G.~M.}\ \bibnamefont
  {Whitesides}}\ and\ \bibinfo {author} {\bibfnamefont {D.~J.}\ \bibnamefont
  {Lipomi}},\ }\href@noop {} {\bibfield  {journal} {\bibinfo  {journal}
  {{F}araday {D}iscuss.}\ }\textbf {\bibinfo {volume} {143}},\ \bibinfo {pages}
  {373} (\bibinfo {year} {2009})}\BibitemShut {NoStop}%
\bibitem [{\citenamefont {Srivastava}\ \emph {et~al.}(2010)\citenamefont
  {Srivastava}, \citenamefont {Santos}, \citenamefont {Critchley},
  \citenamefont {Kim}, \citenamefont {Podsiadlo}, \citenamefont {Sun},
  \citenamefont {Lee}, \citenamefont {Xu}, \citenamefont {Lilly}, \citenamefont
  {Glotzer},\ and\ \citenamefont {Kotov}}]{sri10}%
  \BibitemOpen
  \bibfield  {author} {\bibinfo {author} {\bibfnamefont {S.}~\bibnamefont
  {Srivastava}}, \bibinfo {author} {\bibfnamefont {A.}~\bibnamefont {Santos}},
  \bibinfo {author} {\bibfnamefont {K.}~\bibnamefont {Critchley}}, \bibinfo
  {author} {\bibfnamefont {K.-S.}\ \bibnamefont {Kim}}, \bibinfo {author}
  {\bibfnamefont {P.}~\bibnamefont {Podsiadlo}}, \bibinfo {author}
  {\bibfnamefont {K.}~\bibnamefont {Sun}}, \bibinfo {author} {\bibfnamefont
  {J.}~\bibnamefont {Lee}}, \bibinfo {author} {\bibfnamefont {C.}~\bibnamefont
  {Xu}}, \bibinfo {author} {\bibfnamefont {G.~D.}\ \bibnamefont {Lilly}},
  \bibinfo {author} {\bibfnamefont {S.~C.}\ \bibnamefont {Glotzer}}, \ and\
  \bibinfo {author} {\bibfnamefont {N.~A.}\ \bibnamefont {Kotov}},\ }\href@noop
  {} {\bibfield  {journal} {\bibinfo  {journal} {{S}cience}\ }\textbf {\bibinfo
  {volume} {327}},\ \bibinfo {pages} {1355} (\bibinfo {year}
  {2010})}\BibitemShut {NoStop}%
\bibitem [{\citenamefont {Wang}\ and\ \citenamefont {Zhou}(2010)}]{wang10}%
  \BibitemOpen
  \bibfield  {author} {\bibinfo {author} {\bibfnamefont {Y.}~\bibnamefont
  {Wang}}\ and\ \bibinfo {author} {\bibfnamefont {W.}~\bibnamefont {Zhou}},\
  }\href@noop {} {\bibfield  {journal} {\bibinfo  {journal} {{J}ournal of
  {N}anoscience {A}nd {N}anotechnology}\ }\textbf {\bibinfo {volume} {10}},\
  \bibinfo {pages} {1563} (\bibinfo {year} {2010})}\BibitemShut {NoStop}%
\bibitem [{\citenamefont {Grzelczak}\ \emph {et~al.}(2010)\citenamefont
  {Grzelczak}, \citenamefont {Vermant}, \citenamefont {Furst},\ and\
  \citenamefont {Liz-Marzan}}]{grzel10}%
  \BibitemOpen
  \bibfield  {author} {\bibinfo {author} {\bibfnamefont {M.}~\bibnamefont
  {Grzelczak}}, \bibinfo {author} {\bibfnamefont {J.}~\bibnamefont {Vermant}},
  \bibinfo {author} {\bibfnamefont {E.~M.}\ \bibnamefont {Furst}}, \ and\
  \bibinfo {author} {\bibfnamefont {L.~M.}\ \bibnamefont {Liz-Marzan}},\
  }\href@noop {} {\bibfield  {journal} {\bibinfo  {journal} {{ACS} {N}ano}\
  }\textbf {\bibinfo {volume} {4}},\ \bibinfo {pages} {3591} (\bibinfo {year}
  {2010})}\BibitemShut {NoStop}%
\bibitem [{\citenamefont {Glotzer}\ and\ \citenamefont
  {Solomon}(2007)}]{sol07}%
  \BibitemOpen
  \bibfield  {author} {\bibinfo {author} {\bibfnamefont {S.~C.}\ \bibnamefont
  {Glotzer}}\ and\ \bibinfo {author} {\bibfnamefont {M.~J.}\ \bibnamefont
  {Solomon}},\ }\href@noop {} {\bibfield  {journal} {\bibinfo  {journal}
  {Nature Materials}\ }\textbf {\bibinfo {volume} {6}},\ \bibinfo {pages} {557}
  (\bibinfo {year} {2007})}\BibitemShut {NoStop}%
\bibitem [{\citenamefont {Hong}\ \emph {et~al.}(2006)\citenamefont {Hong},
  \citenamefont {Cacciuto}, \citenamefont {Luijten},\ and\ \citenamefont
  {Granick}}]{hong06}%
  \BibitemOpen
  \bibfield  {author} {\bibinfo {author} {\bibfnamefont {L.}~\bibnamefont
  {Hong}}, \bibinfo {author} {\bibfnamefont {A.}~\bibnamefont {Cacciuto}},
  \bibinfo {author} {\bibfnamefont {E.}~\bibnamefont {Luijten}}, \ and\
  \bibinfo {author} {\bibfnamefont {S.}~\bibnamefont {Granick}},\ }\href@noop
  {} {\bibfield  {journal} {\bibinfo  {journal} {{N}ano {L}ett.}\ }\textbf
  {\bibinfo {volume} {6}},\ \bibinfo {pages} {2510} (\bibinfo {year}
  {2006})}\BibitemShut {NoStop}%
\bibitem [{\citenamefont {Douglas}\ \emph {et~al.}(2009)\citenamefont
  {Douglas}, \citenamefont {Dietz}, \citenamefont {Liedl}, \citenamefont
  {Hogberg}, \citenamefont {Graf},\ and\ \citenamefont {Shih}}]{doug09}%
  \BibitemOpen
  \bibfield  {author} {\bibinfo {author} {\bibfnamefont {S.~M.}\ \bibnamefont
  {Douglas}}, \bibinfo {author} {\bibfnamefont {H.}~\bibnamefont {Dietz}},
  \bibinfo {author} {\bibfnamefont {T.}~\bibnamefont {Liedl}}, \bibinfo
  {author} {\bibfnamefont {B.}~\bibnamefont {Hogberg}}, \bibinfo {author}
  {\bibfnamefont {F.}~\bibnamefont {Graf}}, \ and\ \bibinfo {author}
  {\bibfnamefont {W.~M.}\ \bibnamefont {Shih}},\ }\href@noop {} {\bibfield
  {journal} {\bibinfo  {journal} {Nature}\ }\textbf {\bibinfo {volume} {459}},\
  \bibinfo {pages} {414} (\bibinfo {year} {2009})}\BibitemShut {NoStop}%
\bibitem [{\citenamefont {Sacanna}\ \emph {et~al.}(2010)\citenamefont
  {Sacanna}, \citenamefont {Irvine}, \citenamefont {Chaikin},\ and\
  \citenamefont {Pine}}]{sac10}%
  \BibitemOpen
  \bibfield  {author} {\bibinfo {author} {\bibfnamefont {S.}~\bibnamefont
  {Sacanna}}, \bibinfo {author} {\bibfnamefont {W.~T.~M.}\ \bibnamefont
  {Irvine}}, \bibinfo {author} {\bibfnamefont {P.~M.}\ \bibnamefont {Chaikin}},
  \ and\ \bibinfo {author} {\bibfnamefont {D.~J.}\ \bibnamefont {Pine}},\
  }\href@noop {} {\bibfield  {journal} {\bibinfo  {journal} {{N}ature}\
  }\textbf {\bibinfo {volume} {464}},\ \bibinfo {pages} {575} (\bibinfo {year}
  {2010})}\BibitemShut {NoStop}%
\bibitem [{\citenamefont {Bray}(1994)}]{bray}%
  \BibitemOpen
  \bibfield  {author} {\bibinfo {author} {\bibfnamefont {A.}~\bibnamefont
  {Bray}},\ }\href@noop {} {\bibfield  {journal} {\bibinfo  {journal} {Adv
  Phys}\ }\textbf {\bibinfo {volume} {43}},\ \bibinfo {pages} {357} (\bibinfo
  {year} {1994})},\ \bibinfo {note} {advances In Physics}\BibitemShut {NoStop}%
\bibitem [{\citenamefont {Slabinski}\ \emph {et~al.}(2007)\citenamefont
  {Slabinski}, \citenamefont {Jaroszewski}, \citenamefont {Rodrigues},
  \citenamefont {Rychlewski}, \citenamefont {Wilson}, \citenamefont {Lesley},\
  and\ \citenamefont {Godzik}}]{slab07}%
  \BibitemOpen
  \bibfield  {author} {\bibinfo {author} {\bibfnamefont {L.}~\bibnamefont
  {Slabinski}}, \bibinfo {author} {\bibfnamefont {L.}~\bibnamefont
  {Jaroszewski}}, \bibinfo {author} {\bibfnamefont {A.~P.~C.}\ \bibnamefont
  {Rodrigues}}, \bibinfo {author} {\bibfnamefont {L.}~\bibnamefont
  {Rychlewski}}, \bibinfo {author} {\bibfnamefont {I.~A.}\ \bibnamefont
  {Wilson}}, \bibinfo {author} {\bibfnamefont {S.~A.}\ \bibnamefont {Lesley}},
  \ and\ \bibinfo {author} {\bibfnamefont {A.}~\bibnamefont {Godzik}},\
  }\href@noop {} {\bibfield  {journal} {\bibinfo  {journal} {{P}rotein {S}ci.}\
  }\textbf {\bibinfo {volume} {16}},\ \bibinfo {pages} {2472} (\bibinfo {year}
  {2007})}\BibitemShut {NoStop}%
\bibitem [{\citenamefont {Leng}\ and\ \citenamefont {Salmon}(2009)}]{leng09}%
  \BibitemOpen
  \bibfield  {author} {\bibinfo {author} {\bibfnamefont {J.}~\bibnamefont
  {Leng}}\ and\ \bibinfo {author} {\bibfnamefont {J.-B.}\ \bibnamefont
  {Salmon}},\ }\href@noop {} {\bibfield  {journal} {\bibinfo  {journal} {{L}ab
  {C}hip}\ }\textbf {\bibinfo {volume} {9}},\ \bibinfo {pages} {24} (\bibinfo
  {year} {2009})}\BibitemShut {NoStop}%
\bibitem [{\citenamefont {Lu}\ \emph {et~al.}(2006)\citenamefont {Lu},
  \citenamefont {Conrad}, \citenamefont {Wyss}, \citenamefont {Schofield},\
  and\ \citenamefont {Weitz}}]{lu06}%
  \BibitemOpen
  \bibfield  {author} {\bibinfo {author} {\bibfnamefont {P.~J.}\ \bibnamefont
  {Lu}}, \bibinfo {author} {\bibfnamefont {J.~C.}\ \bibnamefont {Conrad}},
  \bibinfo {author} {\bibfnamefont {H.~M.}\ \bibnamefont {Wyss}}, \bibinfo
  {author} {\bibfnamefont {A.~B.}\ \bibnamefont {Schofield}}, \ and\ \bibinfo
  {author} {\bibfnamefont {D.~A.}\ \bibnamefont {Weitz}},\ }\href@noop {}
  {\bibfield  {journal} {\bibinfo  {journal} {{P}hys. {R}ev. {L}ett.}\ }\textbf
  {\bibinfo {volume} {96}},\ \bibinfo {pages} {028306} (\bibinfo {year}
  {2006})}\BibitemShut {NoStop}%
\bibitem [{\citenamefont {Chayen}(2004)}]{Chayen2004}%
  \BibitemOpen
  \bibfield  {author} {\bibinfo {author} {\bibfnamefont {N.~E.}\ \bibnamefont
  {Chayen}},\ }\href {\doibase DOI: 10.1016/j.sbi.2004.08.002} {\bibfield
  {journal} {\bibinfo  {journal} {Curr. Opinion Struct. Biol.}\ }\textbf
  {\bibinfo {volume} {14}},\ \bibinfo {pages} {577 } (\bibinfo {year}
  {2004})}\BibitemShut {NoStop}%
\bibitem [{\citenamefont {Vlasov}\ \emph {et~al.}(2001)\citenamefont {Vlasov},
  \citenamefont {Bo}, \citenamefont {Sturm},\ and\ \citenamefont
  {Norris}}]{vlasov2001}%
  \BibitemOpen
  \bibfield  {author} {\bibinfo {author} {\bibfnamefont {Y.~A.}\ \bibnamefont
  {Vlasov}}, \bibinfo {author} {\bibfnamefont {X.-Z.}\ \bibnamefont {Bo}},
  \bibinfo {author} {\bibfnamefont {J.~C.}\ \bibnamefont {Sturm}}, \ and\
  \bibinfo {author} {\bibfnamefont {D.~J.}\ \bibnamefont {Norris}},\ }\href
  {http://dx.doi.org/10.1038/35104529} {\bibfield  {journal} {\bibinfo
  {journal} {Nature}\ }\textbf {\bibinfo {volume} {414}},\ \bibinfo {pages}
  {289} (\bibinfo {year} {2001})}\BibitemShut {NoStop}%
\bibitem [{\citenamefont {Hynninen}\ \emph {et~al.}(2007)\citenamefont
  {Hynninen}, \citenamefont {Thijssen}, \citenamefont {Vermolen}, \citenamefont
  {Dijkstra},\ and\ \citenamefont {Van~Blaaderen}}]{hynn2007}%
  \BibitemOpen
  \bibfield  {author} {\bibinfo {author} {\bibfnamefont {A.-P.}\ \bibnamefont
  {Hynninen}}, \bibinfo {author} {\bibfnamefont {J.~H.~J.}\ \bibnamefont
  {Thijssen}}, \bibinfo {author} {\bibfnamefont {E.~C.~M.}\ \bibnamefont
  {Vermolen}}, \bibinfo {author} {\bibfnamefont {M.}~\bibnamefont {Dijkstra}},
  \ and\ \bibinfo {author} {\bibfnamefont {A.}~\bibnamefont {Van~Blaaderen}},\
  }\href {\doibase DOI 10.1038/nmat1841} {\bibfield  {journal} {\bibinfo
  {journal} {Nature Mat.}\ }\textbf {\bibinfo {volume} {6}},\ \bibinfo {pages}
  {202} (\bibinfo {year} {2007})}\BibitemShut {NoStop}%
\bibitem [{\citenamefont {{ten Wolde}}\ and\ \citenamefont
  {Frenkel}(1997)}]{frenkel97}%
  \BibitemOpen
  \bibfield  {author} {\bibinfo {author} {\bibfnamefont {P.}~\bibnamefont {{ten
  Wolde}}}\ and\ \bibinfo {author} {\bibfnamefont {D.}~\bibnamefont
  {Frenkel}},\ }\href@noop {} {\bibfield  {journal} {\bibinfo  {journal}
  {Science}\ }\textbf {\bibinfo {volume} {277}},\ \bibinfo {pages} {1975}
  (\bibinfo {year} {1997})}\BibitemShut {NoStop}%
\bibitem [{\citenamefont {Likos}(2001)}]{likos01}%
  \BibitemOpen
  \bibfield  {author} {\bibinfo {author} {\bibfnamefont {C.~N.}\ \bibnamefont
  {Likos}},\ }\href@noop {} {\bibfield  {journal} {\bibinfo  {journal} {Phys.
  Rep.}\ }\textbf {\bibinfo {volume} {348}},\ \bibinfo {pages} {267} (\bibinfo
  {year} {2001})}\BibitemShut {NoStop}%
\bibitem [{\citenamefont {Foffi}\ \emph {et~al.}(2002)\citenamefont {Foffi},
  \citenamefont {McCullagh}, \citenamefont {Lawlor}, \citenamefont
  {Zaccarelli},\ and\ \citenamefont {Dawson}}]{foffi02}%
  \BibitemOpen
  \bibfield  {author} {\bibinfo {author} {\bibfnamefont {G.}~\bibnamefont
  {Foffi}}, \bibinfo {author} {\bibfnamefont {G.~D.}\ \bibnamefont
  {McCullagh}}, \bibinfo {author} {\bibfnamefont {A.}~\bibnamefont {Lawlor}},
  \bibinfo {author} {\bibfnamefont {E.}~\bibnamefont {Zaccarelli}}, \ and\
  \bibinfo {author} {\bibfnamefont {K.~A.}\ \bibnamefont {Dawson}},\
  }\href@noop {} {\bibfield  {journal} {\bibinfo  {journal} {{P}hys. {R}ev.
  {E}}\ }\textbf {\bibinfo {volume} {65}},\ \bibinfo {pages} {031407} (\bibinfo
  {year} {2002})}\BibitemShut {NoStop}%
\bibitem [{\citenamefont {Dijkstra}(2002)}]{dijkstra02}%
  \BibitemOpen
  \bibfield  {author} {\bibinfo {author} {\bibfnamefont {M.}~\bibnamefont
  {Dijkstra}},\ }\href@noop {} {\bibfield  {journal} {\bibinfo  {journal}
  {{P}hys. {R}ev. {E}}\ }\textbf {\bibinfo {volume} {66}},\ \bibinfo {pages}
  {021402} (\bibinfo {year} {2002})}\BibitemShut {NoStop}%
\bibitem [{\citenamefont {Tavares}\ and\ \citenamefont
  {Prausnitz}(2004)}]{tava04}%
  \BibitemOpen
  \bibfield  {author} {\bibinfo {author} {\bibfnamefont {F.~W.}\ \bibnamefont
  {Tavares}}\ and\ \bibinfo {author} {\bibfnamefont {J.~M.}\ \bibnamefont
  {Prausnitz}},\ }\href@noop {} {\bibfield  {journal} {\bibinfo  {journal}
  {{C}olloid {P}olym. {S}ci.}\ }\textbf {\bibinfo {volume} {282}},\ \bibinfo
  {pages} {620} (\bibinfo {year} {2004})}\BibitemShut {NoStop}%
\bibitem [{\citenamefont {Charbonneau}\ and\ \citenamefont
  {Reichman}(2007)}]{charb07}%
  \BibitemOpen
  \bibfield  {author} {\bibinfo {author} {\bibfnamefont {P.}~\bibnamefont
  {Charbonneau}}\ and\ \bibinfo {author} {\bibfnamefont {D.~R.}\ \bibnamefont
  {Reichman}},\ }\href@noop {} {\bibfield  {journal} {\bibinfo  {journal}
  {{P}hys. {R}ev. {E}}\ }\textbf {\bibinfo {volume} {75}},\ \bibinfo {pages}
  {011507} (\bibinfo {year} {2007})}\BibitemShut {NoStop}%
\bibitem [{\citenamefont {Fortini}\ \emph {et~al.}(2008)\citenamefont
  {Fortini}, \citenamefont {Sanz},\ and\ \citenamefont {Dijkstra}}]{fortini08}%
  \BibitemOpen
  \bibfield  {author} {\bibinfo {author} {\bibfnamefont {A.}~\bibnamefont
  {Fortini}}, \bibinfo {author} {\bibfnamefont {E.}~\bibnamefont {Sanz}}, \
  and\ \bibinfo {author} {\bibfnamefont {M.}~\bibnamefont {Dijkstra}},\ }\href
  {\doibase DOI 10.1103/PhysRevE.78.041402} {\bibfield  {journal} {\bibinfo
  {journal} {{P}hys. {R}ev. {E}}\ }\textbf {\bibinfo {volume} {78}},\ \bibinfo
  {pages} {041402} (\bibinfo {year} {2008})}\BibitemShut {NoStop}%
\bibitem [{\citenamefont {Whitelam}\ \emph {et~al.}(2009)\citenamefont
  {Whitelam}, \citenamefont {Feng}, \citenamefont {Hagan},\ and\ \citenamefont
  {Geissler}}]{whitelam09}%
  \BibitemOpen
  \bibfield  {author} {\bibinfo {author} {\bibfnamefont {S.}~\bibnamefont
  {Whitelam}}, \bibinfo {author} {\bibfnamefont {E.~H.}\ \bibnamefont {Feng}},
  \bibinfo {author} {\bibfnamefont {M.~F.}\ \bibnamefont {Hagan}}, \ and\
  \bibinfo {author} {\bibfnamefont {P.~L.}\ \bibnamefont {Geissler}},\
  }\href@noop {} {\bibfield  {journal} {\bibinfo  {journal} {{S}oft {M}atter}\
  }\textbf {\bibinfo {volume} {5}},\ \bibinfo {pages} {1251} (\bibinfo {year}
  {2009})}\BibitemShut {NoStop}%
\bibitem [{\citenamefont {Liu}\ \emph {et~al.}(2005)\citenamefont {Liu},
  \citenamefont {Garde},\ and\ \citenamefont {Kumar}}]{liu05}%
  \BibitemOpen
  \bibfield  {author} {\bibinfo {author} {\bibfnamefont {H.}~\bibnamefont
  {Liu}}, \bibinfo {author} {\bibfnamefont {S.}~\bibnamefont {Garde}}, \ and\
  \bibinfo {author} {\bibfnamefont {S.}~\bibnamefont {Kumar}},\ }\href@noop {}
  {\bibfield  {journal} {\bibinfo  {journal} {{J}. {C}hem. {P}hys.}\ }\textbf
  {\bibinfo {volume} {123}},\ \bibinfo {pages} {174505} (\bibinfo {year}
  {2005})}\BibitemShut {NoStop}%
\bibitem [{\citenamefont {Honeycutt}\ and\ \citenamefont
  {Andersen}(1987)}]{hon87}%
  \BibitemOpen
  \bibfield  {author} {\bibinfo {author} {\bibfnamefont {J.~D.}\ \bibnamefont
  {Honeycutt}}\ and\ \bibinfo {author} {\bibfnamefont {H.~C.}\ \bibnamefont
  {Andersen}},\ }\href@noop {} {\bibfield  {journal} {\bibinfo  {journal} {{J}.
  {P}hys. {C}hem.}\ }\textbf {\bibinfo {volume} {91}},\ \bibinfo {pages} {4950}
  (\bibinfo {year} {1987})}\BibitemShut {NoStop}%
\bibitem [{\citenamefont {Jack}\ \emph {et~al.}(2007)\citenamefont {Jack},
  \citenamefont {Hagan},\ and\ \citenamefont {Chandler}}]{jack07}%
  \BibitemOpen
  \bibfield  {author} {\bibinfo {author} {\bibfnamefont {R.~L.}\ \bibnamefont
  {Jack}}, \bibinfo {author} {\bibfnamefont {M.~F.}\ \bibnamefont {Hagan}}, \
  and\ \bibinfo {author} {\bibfnamefont {D.}~\bibnamefont {Chandler}},\
  }\href@noop {} {\bibfield  {journal} {\bibinfo  {journal} {{P}hys. {R}ev.
  {E}}\ }\textbf {\bibinfo {volume} {76}},\ \bibinfo {pages} {021119} (\bibinfo
  {year} {2007})}\BibitemShut {NoStop}%
\bibitem [{\citenamefont {Asherie}\ \emph {et~al.}(1996)\citenamefont
  {Asherie}, \citenamefont {Lomakin},\ and\ \citenamefont {Benedek}}]{ash96}%
  \BibitemOpen
  \bibfield  {author} {\bibinfo {author} {\bibfnamefont {N.}~\bibnamefont
  {Asherie}}, \bibinfo {author} {\bibfnamefont {A.}~\bibnamefont {Lomakin}}, \
  and\ \bibinfo {author} {\bibfnamefont {G.~B.}\ \bibnamefont {Benedek}},\
  }\href@noop {} {\bibfield  {journal} {\bibinfo  {journal} {Phys. Rev. Lett.}\
  }\textbf {\bibinfo {volume} {77}},\ \bibinfo {pages} {4832} (\bibinfo {year}
  {1996})}\BibitemShut {NoStop}%
\bibitem [{\citenamefont {Duda}(2009)}]{duda09}%
  \BibitemOpen
  \bibfield  {author} {\bibinfo {author} {\bibfnamefont {Y.}~\bibnamefont
  {Duda}},\ }\href@noop {} {\bibfield  {journal} {\bibinfo  {journal} {J. Chem.
  Phys.}\ }\textbf {\bibinfo {volume} {130}},\ \bibinfo {pages} {116101}
  (\bibinfo {year} {2009})}\BibitemShut {NoStop}%
\bibitem [{Note1()}]{Note1}%
  \BibitemOpen
  \bibinfo {note} {Strictly, $\nabla _i E$ is ill-defined, since $E$ is not a
  continuous function of the particle coordinates. For the purposes of (\ref
  {equ:lang}) we imagine regularising the square-well potential by taking a
  limiting case of a smooth but steep potential. In practice, we integrate this
  equation using a Monte Carlo scheme with a finite time step $\tau _0$, which
  avoids the need for any explicit regularisation.}\BibitemShut {Stop}%
\bibitem [{Note2()}]{Note2}%
  \BibitemOpen
  \bibinfo {note} {An alternative to our MC scheme would be to use Brownian
  dynamics to simulate this system: in the limit of small time step $\tau _0$
  then both Brownian and MC dynamics are equivalent. One reason to prefer the
  MC in this study is that the fluctuation-dissipation theorems described in
  the following sections hold exactly for equilibrated systems with MC
  dynamics, even when the time step $\tau _0$ is finite. (This is not the case
  when using Brownian dynamics.)}\BibitemShut {NoStop}%
\bibitem [{\citenamefont {Whitesides}\ and\ \citenamefont
  {Boncheva}(2002)}]{white02}%
  \BibitemOpen
  \bibfield  {author} {\bibinfo {author} {\bibfnamefont {G.~M.}\ \bibnamefont
  {Whitesides}}\ and\ \bibinfo {author} {\bibfnamefont {M.}~\bibnamefont
  {Boncheva}},\ }\href@noop {} {\bibfield  {journal} {\bibinfo  {journal}
  {{Proc. Natl. Acad. Sci. U. S. A.}}\ }\textbf {\bibinfo {volume} {99}},\
  \bibinfo {pages} {4769} (\bibinfo {year} {2002})}\BibitemShut {NoStop}%
\bibitem [{\citenamefont {Cugliandolo}\ \emph {et~al.}(1997)\citenamefont
  {Cugliandolo}, \citenamefont {Kurchan},\ and\ \citenamefont
  {Peliti}}]{cuglia97}%
  \BibitemOpen
  \bibfield  {author} {\bibinfo {author} {\bibfnamefont {L.~F.}\ \bibnamefont
  {Cugliandolo}}, \bibinfo {author} {\bibfnamefont {J.}~\bibnamefont
  {Kurchan}}, \ and\ \bibinfo {author} {\bibfnamefont {L.}~\bibnamefont
  {Peliti}},\ }\href@noop {} {\bibfield  {journal} {\bibinfo  {journal}
  {{P}hys. {R}ev. {E}}\ }\textbf {\bibinfo {volume} {55}},\ \bibinfo {pages}
  {3898} (\bibinfo {year} {1997})}\BibitemShut {NoStop}%
\bibitem [{\citenamefont {Crisanti}\ and\ \citenamefont
  {Ritort}(2003)}]{crisanti03}%
  \BibitemOpen
  \bibfield  {author} {\bibinfo {author} {\bibfnamefont {A.}~\bibnamefont
  {Crisanti}}\ and\ \bibinfo {author} {\bibfnamefont {F.}~\bibnamefont
  {Ritort}},\ }\href@noop {} {\bibfield  {journal} {\bibinfo  {journal} {{J}.
  {P}hys. {A}: {M}ath. {G}en.}\ }\textbf {\bibinfo {volume} {36}},\ \bibinfo
  {pages} {R181} (\bibinfo {year} {2003})}\BibitemShut {NoStop}%
\bibitem [{\citenamefont {Kurchan}(2005)}]{kurchan05}%
  \BibitemOpen
  \bibfield  {author} {\bibinfo {author} {\bibfnamefont {J.}~\bibnamefont
  {Kurchan}},\ }\href@noop {} {\bibfield  {journal} {\bibinfo  {journal}
  {{N}ature}\ }\textbf {\bibinfo {volume} {433}},\ \bibinfo {pages} {222}
  (\bibinfo {year} {2005})}\BibitemShut {NoStop}%
\bibitem [{\citenamefont {Baiesi}\ \emph {et~al.}(2009)\citenamefont {Baiesi},
  \citenamefont {Maes},\ and\ \citenamefont {Wynants}}]{baiesi09}%
  \BibitemOpen
  \bibfield  {author} {\bibinfo {author} {\bibfnamefont {M.}~\bibnamefont
  {Baiesi}}, \bibinfo {author} {\bibfnamefont {C.}~\bibnamefont {Maes}}, \ and\
  \bibinfo {author} {\bibfnamefont {B.}~\bibnamefont {Wynants}},\ }\href
  {\doibase DOI 10.1103/PhysRevLett.103.010602} {\bibfield  {journal} {\bibinfo
   {journal} {Physical Review Letters}\ }\textbf {\bibinfo {volume} {103}},\
  \bibinfo {pages} {010602} (\bibinfo {year} {2009})}\BibitemShut {NoStop}%
\bibitem [{\citenamefont {Seifert}\ and\ \citenamefont
  {Speck}(2010)}]{seifert10}%
  \BibitemOpen
  \bibfield  {author} {\bibinfo {author} {\bibfnamefont {U.}~\bibnamefont
  {Seifert}}\ and\ \bibinfo {author} {\bibfnamefont {T.}~\bibnamefont
  {Speck}},\ }\href {\doibase DOI 10.1209/0295-5075/89/10007} {\bibfield
  {journal} {\bibinfo  {journal} {EPL}\ }\textbf {\bibinfo {volume} {89}},\
  \bibinfo {pages} {10007} (\bibinfo {year} {2010})}\BibitemShut {NoStop}%
\bibitem [{\citenamefont {Russo}\ and\ \citenamefont
  {Sciortino}(2010)}]{russo10}%
  \BibitemOpen
  \bibfield  {author} {\bibinfo {author} {\bibfnamefont {J.}~\bibnamefont
  {Russo}}\ and\ \bibinfo {author} {\bibfnamefont {F.}~\bibnamefont
  {Sciortino}},\ }\href@noop {} {\bibfield  {journal} {\bibinfo  {journal}
  {{P}hys. {R}ev. {L}ett.}\ }\textbf {\bibinfo {volume} {104}},\ \bibinfo
  {pages} {195701} (\bibinfo {year} {2010})}\BibitemShut {NoStop}%
\bibitem [{\citenamefont {Jabbari-Farouji}\ \emph {et~al.}(2007)\citenamefont
  {Jabbari-Farouji}, \citenamefont {Mizuno}, \citenamefont {Atakhorrami},
  \citenamefont {MacKintosh}, \citenamefont {Christoph F.~Schmidt},
  \citenamefont {Wegdam},\ and\ \citenamefont {Bonn}}]{bonn07a}%
  \BibitemOpen
  \bibfield  {author} {\bibinfo {author} {\bibfnamefont {S.}~\bibnamefont
  {Jabbari-Farouji}}, \bibinfo {author} {\bibfnamefont {D.}~\bibnamefont
  {Mizuno}}, \bibinfo {author} {\bibfnamefont {M.}~\bibnamefont {Atakhorrami}},
  \bibinfo {author} {\bibfnamefont {F.~C.}\ \bibnamefont {MacKintosh}},
  \bibinfo {author} {\bibfnamefont {E.~E.}\ \bibnamefont {Christoph
  F.~Schmidt}}, \bibinfo {author} {\bibfnamefont {G.~H.}\ \bibnamefont
  {Wegdam}}, \ and\ \bibinfo {author} {\bibfnamefont {D.}~\bibnamefont
  {Bonn}},\ }\href@noop {} {\bibfield  {journal} {\bibinfo  {journal} {{P}hys.
  {R}ev. {L}ett.}\ }\textbf {\bibinfo {volume} {98}},\ \bibinfo {pages}
  {108302} (\bibinfo {year} {2007})}\BibitemShut {NoStop}%
\bibitem [{\citenamefont {Maggi}\ \emph {et~al.}(2010)\citenamefont {Maggi},
  \citenamefont {Leonardo}, \citenamefont {Dyre},\ and\ \citenamefont
  {Ruocco}}]{ruocco10}%
  \BibitemOpen
  \bibfield  {author} {\bibinfo {author} {\bibfnamefont {C.}~\bibnamefont
  {Maggi}}, \bibinfo {author} {\bibfnamefont {R.~D.}\ \bibnamefont {Leonardo}},
  \bibinfo {author} {\bibfnamefont {J.~C.}\ \bibnamefont {Dyre}}, \ and\
  \bibinfo {author} {\bibfnamefont {G.}~\bibnamefont {Ruocco}},\ }\href@noop {}
  {\bibfield  {journal} {\bibinfo  {journal} {{P}hys. {R}ev. {B}}\ }\textbf
  {\bibinfo {volume} {81}},\ \bibinfo {pages} {104201} (\bibinfo {year}
  {2010})}\BibitemShut {NoStop}%
\bibitem [{\citenamefont {Oukris}\ and\ \citenamefont
  {Israeloff}(2010)}]{oukris10}%
  \BibitemOpen
  \bibfield  {author} {\bibinfo {author} {\bibfnamefont {H.}~\bibnamefont
  {Oukris}}\ and\ \bibinfo {author} {\bibfnamefont {N.~E.}\ \bibnamefont
  {Israeloff}},\ }\href@noop {} {\bibfield  {journal} {\bibinfo  {journal}
  {{N}ature {P}hysics}\ }\textbf {\bibinfo {volume} {6}},\ \bibinfo {pages}
  {135} (\bibinfo {year} {2010})}\BibitemShut {NoStop}%
\bibitem [{\citenamefont {Sollich}\ \emph {et~al.}(2002)\citenamefont
  {Sollich}, \citenamefont {Fielding},\ and\ \citenamefont
  {Mayer}}]{sollich02}%
  \BibitemOpen
  \bibfield  {author} {\bibinfo {author} {\bibfnamefont {P.}~\bibnamefont
  {Sollich}}, \bibinfo {author} {\bibfnamefont {S.}~\bibnamefont {Fielding}}, \
  and\ \bibinfo {author} {\bibfnamefont {P.}~\bibnamefont {Mayer}},\
  }\href@noop {} {\bibfield  {journal} {\bibinfo  {journal} {{J}. {P}hys.
  {C}ondens. {M}atter}\ }\textbf {\bibinfo {volume} {14}},\ \bibinfo {pages}
  {1683} (\bibinfo {year} {2002})}\BibitemShut {NoStop}%
\bibitem [{\citenamefont {Jack}\ \emph {et~al.}(2006)\citenamefont {Jack},
  \citenamefont {Berthier},\ and\ \citenamefont {Garrahan}}]{jack-fdt06}%
  \BibitemOpen
  \bibfield  {author} {\bibinfo {author} {\bibfnamefont {R.~L.}\ \bibnamefont
  {Jack}}, \bibinfo {author} {\bibfnamefont {L.}~\bibnamefont {Berthier}}, \
  and\ \bibinfo {author} {\bibfnamefont {J.~P.}\ \bibnamefont {Garrahan}},\
  }\href {\doibase ARTN P12005} {\bibfield  {journal} {\bibinfo  {journal} {J.
  Stat. Mech.}\ ,\ \bibinfo {pages} {P12005}} (\bibinfo {year}
  {2006})}\BibitemShut {NoStop}%
\bibitem [{Note3()}]{Note3}%
  \BibitemOpen
  \bibinfo {note} {Our use of the term `local equilibration' is similar in
  spirit to an analogous condition in non-equilibrium thermodynamics~\cite
  {noneq-thermo-book}, but here we are referring to locality in a region of
  configuration space, and not in a spatially localised region of the
  system.}\BibitemShut {Stop}%
\bibitem [{\citenamefont {Rapaport}(2008)}]{rap08}%
  \BibitemOpen
  \bibfield  {author} {\bibinfo {author} {\bibfnamefont {D.~C.}\ \bibnamefont
  {Rapaport}},\ }\href {\doibase ARTN 186101} {\bibfield  {journal} {\bibinfo
  {journal} {{P}hys. {R}ev. {L}ett.}\ }\textbf {\bibinfo {volume} {101}},\
  \bibinfo {pages} {186101} (\bibinfo {year} {2008})}\BibitemShut {NoStop}%
\bibitem [{\citenamefont {Hagan}\ and\ \citenamefont
  {Chandler}(2006)}]{hagan06}%
  \BibitemOpen
  \bibfield  {author} {\bibinfo {author} {\bibfnamefont {M.}~\bibnamefont
  {Hagan}}\ and\ \bibinfo {author} {\bibfnamefont {D.}~\bibnamefont
  {Chandler}},\ }\href {\doibase DOI 10.1529/biophysj.105.076851} {\bibfield
  {journal} {\bibinfo  {journal} {Biophys. J.}\ }\textbf {\bibinfo {volume}
  {91}},\ \bibinfo {pages} {42} (\bibinfo {year} {2006})}\BibitemShut {NoStop}%
\bibitem [{\citenamefont {de~Groot}\ and\ \citenamefont
  {Mazur}(1984)}]{noneq-thermo-book}%
  \BibitemOpen
  \bibfield  {author} {\bibinfo {author} {\bibfnamefont {S.}~\bibnamefont
  {de~Groot}}\ and\ \bibinfo {author} {\bibfnamefont {P.}~\bibnamefont
  {Mazur}},\ }\href@noop {} {\emph {\bibinfo {title} {Non-equilibrium
  thermodynamics}}}\ (\bibinfo  {publisher} {Dover},\ \bibinfo {address}
  {Mineola NY},\ \bibinfo {year} {1984})\BibitemShut {NoStop}%
\end{thebibliography}%
